\documentclass{elsart4}
\pdfoutput=1
\usepackage[utf8]{inputenc}
\usepackage[T1]{fontenc}
\usepackage{amsmath,amsfonts,amssymb,empheq}
\usepackage{dsfont}
\usepackage[english,francais]{babel}
\usepackage[square,numbers,comma,sort&compress]{natbib}
\usepackage{graphicx}

\def\cal{\mathcal}

\begin{document}
\centerline{Physics}
\begin{frontmatter}

\selectlanguage{english}
\title{Ac Josephson Effect in Topological Josephson Junctions}
\author[addressFR]{Driss M. Badiane},
\ead{driss.badiane@gmail.com}
\author[addressUS]{Leonid I. Glazman},
\ead{leonid.glazman@yale.edu} 
\author[addressFR]{Manuel Houzet},
\ead{manuel.houzet@cea.fr}
\author[addressFR]{Julia S. Meyer},
\ead{julia.meyer@ujf-grenoble.fr}
\address[addressFR]{SPSMS, UMR-E 9001 CEA/UJF-Grenoble 1, INAC, Grenoble, F-38054, France}
\address[addressUS]{Department of Physics and Applied Physics, Yale University, New Haven, Connecticut 06520, USA}

\medskip
\begin{center}
{\small Received *****; accepted after revision +++++}
\end{center}

\begin{abstract}

Topological superconductors admit {zero-energy} Majorana bound states at their boundaries. In this review article, we discuss how to probe these Majorana bound states in Josephson junctions between two topological superconductors. In the absence of an applied bias, the presence of these states gives rise to an Andreev bound state whose energy varies $4\pi$-periodically in the superconducting phase difference. An applied voltage bias leads to a dynamically varying phase according to the Josephson relation. Furthermore, it leads to dynamics of the occupation of the bound state via its non-adiabatic coupling to the continuum. While the Josephson relation suggests a fractional Josephson effect due to the $4\pi$-periodicity of the bound state, its observability relies on the conservation of the occupation of the bound state on the experimentally probed time scale. We study the lifetime of the bound state and identify the time scales it has to be compared to. In particular, we are interested in signatures of the fractional Josephson effect in the Shapiro steps and in current noise measurements. We also discuss manifestations of the zero-energy Majorana states on the dissipative subgap current.

\vskip 0.5\baselineskip

\selectlanguage{francais}
\noindent {\bf R\'{e}sum\'{e}}
\vskip 0.5\baselineskip

\noindent {\bf Effet Josephson alternatif dans les jonctions Josephson topologiques. }
Les supraconducteurs topologiques admettent des fermions de Majorana d'\'{e}nergie nulle \`{a} leurs bords. Dans cet article de revue, 
nous discutons la mani\`{e}re de sonder ces \'{e}tats li\'{e}s de Majorana dans une jonction Josephson entre 
deux supraconducteurs topologiques. En l'absence d'une tension de polarisation, la pr\'{e}sence de ces \'{e}tats donne lieu \`{a} un \'{e}tat li\'{e} 
d'Andreev dont l'\'{e}nergie varie $4\pi$-p\'{e}riodiquement vis-\`{a}-vis de la diff\'{e}rence de phase supraconductrice. 
L'application d'une tension de polarisation induit une variation dynamique de la phase en accord avec la relation Josephson. De plus, elle donne lieu \`{a} une dynamique de l'occupation de l'\'{e}tat li\'{e} \`{a} travers son couplage non-adiabatique avec les \'{e}tats du continuum. Tandis que la relation 
de Josephson sugg\`{e}re un effet Josephson fractionnaire d\^{u} \`{a} la  $4\pi$-p\'{e}riodicit\'{e}, son observabilit\'{e} repose sur la conservation de l'occupation de l'\'{e}tat li\'{e} sur l'\'{e}chelle de temps sond\'{e}e exp\'{e}rimentalement. Nous \'{e}tudions la dur\'{e}e de vie de l'\'{e}tat li\'{e} et identifions les \'{e}chelles de temps auxquelles celle-ci doit \^{e}tre compar\'{e}e. En particulier, nous nous int\'{e}ressons aux signatures de l'effet Josephson fractionnaire 
dans les mesures de marches de Shapiro et du bruit en courant. Nous discutons \'{e}galement les manifestations des \'{e}tats de Majorana \`{a} \'{e}nergie nulle dans le courant dissipatif aux tensions plus petites que le gap supraconducteur.

\selectlanguage{english}
\keyword{Majorana fermions; fractional Josephson effect; topological insulators} \vskip 0.5\baselineskip
\selectlanguage{francais}
\noindent{\small{\it Mots-clés~:} fermions de Majorana~; effet Josephson fractionnaire~; isolants topologiques}}
\end{abstract}
\end{frontmatter}

\selectlanguage{english}

\section{Introduction}

Majorana fermions were introduced in 1937 by E. Majorana as solutions of the relativistic Dirac equation~\cite{MajoranaEttore1937}. These fermions are described by real valued fields and, thus, are their own antiparticles. Even if their existence remains hypothetic in high energy physics, recent developments in condensed matter physics suggest their presence as emergent excitations in solid state devices (for reviews in this rapidly growing research field we refer the reader to~\cite{Alicea2012,rev-flensberg,rev-beenakker}). They have attracted a lot of interest due to their non-Abelian statistics, allowing promising applications in quantum computing~{\cite{Kitaev2003,Bravyi2002}}.

Initial proposals for observing Majorana fermions in solid state devices considered the $\nu=5/2$ fractional quantum Hall effect~{\cite{MooreRead1991}}, {superfluid Helium 3~\cite{Kopnin1991},} or quantum spin systems~\cite{Levitov2001,Kitaev2003}. Recently, a great effort has been put towards the observation of Majorana fermions in systems involving superconductors, both on the theoretical and experimental sides. For instance, Majorana fermions appear as zero energy modes at the boundaries of one-dimensional spinless $p$-wave superconductors~{\cite{Kitaev2001}}. They are also trapped in the vortex cores of two-dimensional chiral $p_x + i p_y$ superconductors~{\cite{Volovik1999,ReadGreen2010,Ivanov2001}}. However, superconductors realizing a spin-triplet $p$-wave pairing are not common in nature, strontium ruthenate ($\rm{Sr_2RuO_4}$) being the only candidate so far (for a review on this compound, see~\cite{Mackenzie2003} and references therein). Another scheme which overcomes this difficulty resides in the possibility to artificially engineer a topological superconductor with three generic ingredients that are experimentally accessible within the current state of the art: the proximity effect in the vicinity of a conventional $s$-wave superconductor, spin-orbit coupling, and time-reversal symmetry breaking. The first proposals in this direction were based on using  three-dimensional topological insulators~\cite{Fu2008} or two-dimensional topological insulator, so-called quantum spin-Hall (QSH) insulators~\cite{Fu2009}. QSH insulators are a new class of insulating materials that admit metallic helical edge states~\cite{Bernevig2006,Liu2008,Hasan2010,Qi2011}. Their existence has been confirmed experimentally in transport measurements on HgTe/CdTe~\cite{Konig2007} and InAs/GaSb~\cite{Knez2011} semiconductor heterostructures. When superconductivity is induced within the helical edge states in proximity with a conventional $s$-wave superconductor, the induced superconductivity is effectively spinless $p$-wave. Later it was realized that topological superconductivity may also be realized in nanowires in the presence of both strong spin-orbit coupling and a Zeeman magnetic field~\cite{Lutchyn2010,Oreg2010} by inducing superconducting correlations with a conventional $s$-wave superconductor. 

Zero-energy Majorana states appearing at the boundary of topological superconductors can be probed in tunneling spectroscopy experiments, where they are expected to give rise to a quantized zero-bias conductance,
$G=2e^2/h$~\cite{Law2009,Fidkowski2012}. Recent experimental findings have reported a zero-bias anomaly in the differential conductance of nanowires with a strong spin-orbit coupling, in the presence of a Zeeman field, and in proximity with a superconductor~\cite{Mourik2012,Das2012}. However, a number of other effects 
such as, e.g., disorder \cite{Liu2012,bagrets2012}, the Kondo effect \cite{Lee2012}, or a spin-split Andreev bound state \cite{Lee2013}
may also produce a zero-bias anomaly and ruling them out completely is difficult. Therefore, to unambiguously show the presence of a Majorana fermion, further experiments are needed.

Another predicted signature of Majorana fermions is the appearance of a fractional Josephson effect in topological Josephson junctions \cite{Kitaev2001,Kwon2004}. In a topological Josephson junction, zero-energy Majorana bound states localized on either side of the junction can form an Andreev bound state whose energy varies $4\pi$-periodically with the phase difference between the two superconductors. If the occupation of this bound state were fixed, the Josephson relation $\dot\varphi=2eV$ would then result in a fractional Josephson effect at half of the \lq\lq usual\rq\rq\ Josephson frequency, $\omega_J/2=eV_{\rm dc}$.\footnote{In the remainder of the paper, we use units with $\hbar = k_B = 1$.}  Measuring the fractional Josephson effect would be an additional probe in favor of the presence of Majorana fermions. It is thus important to establish the conditions for the observability of this effect. 

Different methods can be used to detect the ac Josephson effect. One may measure the so-called Shapiro steps~\cite{Shapiro1963} which appear in the presence of an additional ac bias, when the Josephson frequency matches a multiple of the ac frequency $\Omega$. In the case of the fractional Josephson effect, one  expects Shapiro steps at $eV_{\rm{dc}}=k\Omega$ ($k \in \mathds Z$), which corresponds to the even Shapiro steps only of a conventional Josephson junctions~\cite{Kwon2004,Liang2011,Dominguez2012}. Alternatively, one may measure the Josephson radiation or, equivalently, the current noise spectrum of the junction which displays a peak at the Josephson frequency \cite{Yanson,Yanson2}. In a topological Josephson junction, this peak is expected to appear at $eV_{\rm dc}$, i.e., at half of the conventional Josephson frequency~\cite{Kwon2004,Badiane2011}.

In conventional Josephson junction, the visibility of these two effects is limited by the fluctuations of the superconducting phase difference across the junction, which originate from the external circuit the junction is embedded in~\cite{Ivanchenko1969,Ambegaokar1969,Anderson1969}. In topological Josephson junctions, the situation deserves more care. In addition to phase fluctuations, the dynamics of the occupation of the bound state has to be considered. Its occupation may change either because of inelastic processes \cite{Fu2009}, or because the applied bias itself leads to a dynamic coupling between the bound state and the continuum of states above the superconducting gap. In particular, this intrinsic coupling provides an unavoidable mechanism that alters the fractional Josephson effect~\cite{Badiane2011,Houzet2013}.

In this article, we review the properties of a voltage-biased topological Josephson junction to address the observability of the fractional Josephson effect. The outline is as follows. In section~\ref{sec:section_1}, we review the equilibrium properties of a topological Josephson junction based on the helical edge states of a QSH insulator.
In section~\ref{sec:section_2}, we introduce a phenomenological model that allows us to study the dynamics of the bound state in the presence of an applied voltage. We identify the relevant time scales and, then, discuss the observability of the fractional Josephson effect, both in the Shapiro steps and in the current noise spectrum. In section~\ref{sec:section_3}, we review an alternative description of the system based on multiple Andreev reflections. This allows us to study signatures in the noise spectrum in a wider range of parameters. In section~\ref{sec5}, we compare the two approaches introduced in the previous sections. In section~\ref{sec6}, we show that signatures of the presence of Majorana fermions also appear in the dc current. Finally, section~\ref{sec:section_4} summarizes the results discussed in this paper. 

\section{Andreev bound states and Majorana fermions in topological Josephson junctions}\label{sec:section_1}

\begin{figure}[ht!]
\begin{center}
\includegraphics[scale=0.35]{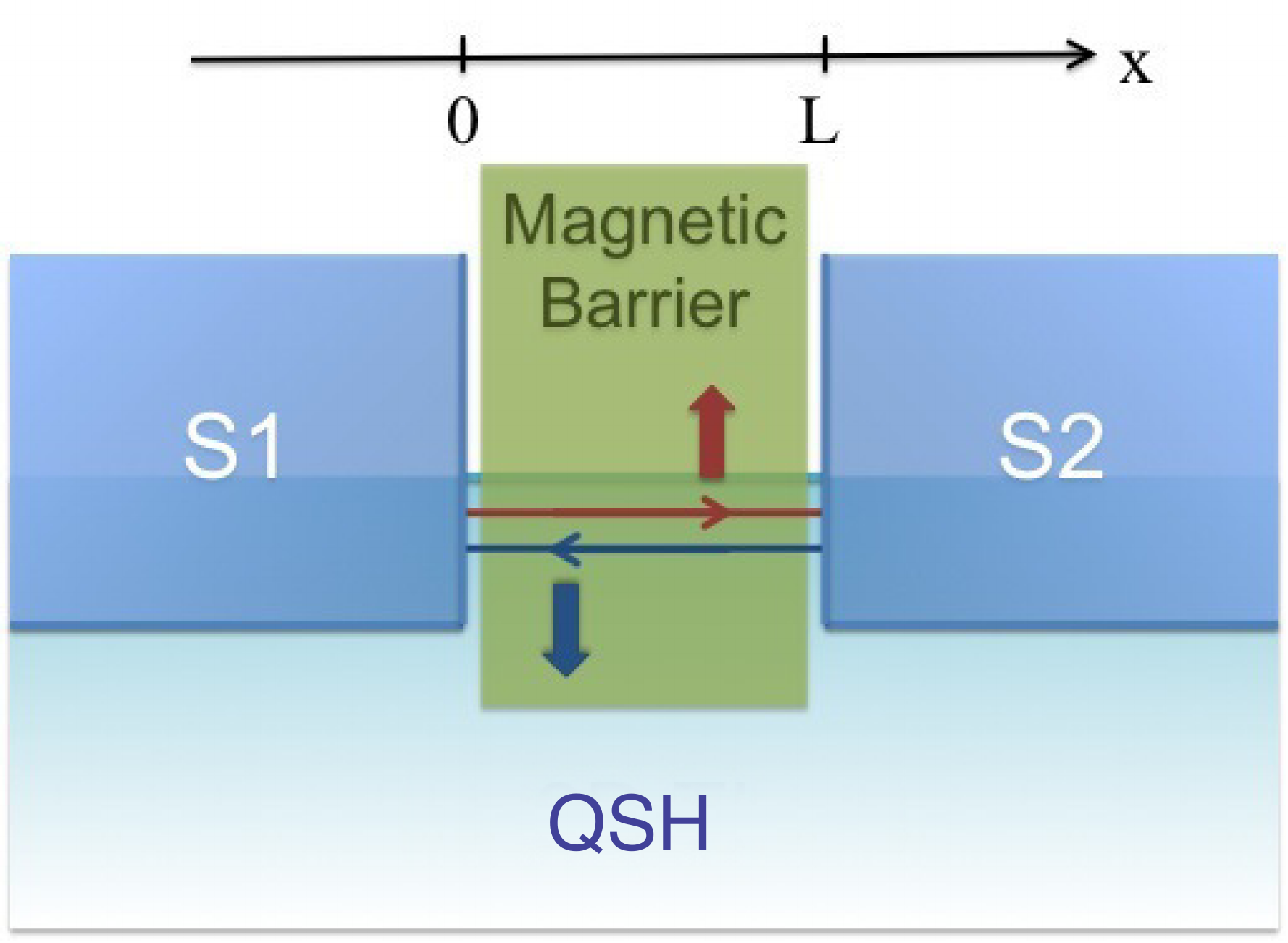}
\caption{Schematic view of the topological Josephson junction. Superconductivity is induced in the helical edge states of the QSH underneath the two superconducting contacts. A transverse magnetic field in the junctions leads to spin-flip back-scattering.}
\label{fig:Fig01}
\end{center}
\end{figure}

To set the stage, let us first discuss a concrete model for a topological Josephson junction \cite{Fu2009} and review its equilibrium properties. In particular, we will take the helical edge states of a QSH insulator as a starting point, cf. Fig.~\ref{fig:Fig01}. Introducing Nambu space, in order to incorporate superconducting correlations later, they are described by the Hamiltonian
\begin{equation}\label{eq:H0}
\mathcal{H}_K=vp_x\sigma_z\tau_z,
\end{equation}
where $v$ is the Fermi velocity and $p_x$ is the momentum operator. Furthermore, $\sigma_i,\tau_j$ ($i,j=x,y,z$) are Pauli matrices acting on the spin and Nambu spaces, respectively. For simplicity, we set the chemical potential $\mu$ to zero, that is to the Dirac point where the helical bands cross.

By attaching superconducting leads, superconductivity may be induced in these helical edge states, underneath the superconducting leads. The proximity induced gap will be denoted $\Delta$. We consider two leads at $x<0$ and $x>L$, respectively. The induced superconducting correlations are then described by the Hamiltonian
\begin{equation}\label{eq:HDelta}
\mathcal{H}_\Delta=\Delta(x)e^{i\phi(x)\tau_z}\tau_x,
\end{equation}
where $\Delta(x)=\Delta\left[\theta(-x)+\theta(x-L)\right]$ and $\phi(x)=\varphi\left[\theta(-x)-\theta(x-L)\right]/2$ with $\pm\varphi/2$ being the superconducting phase of the left and right lead, respectively.

Due to the helical nature of the edge states, a potential barrier does not lead to backscattering. However, a transverse magnetic field allows for spin-flip scattering, thus coupling left- and right-movers. Therefore we include a magnetic barrier, which may be realized by depositing a ferromagnetic insulator. It is described by the Hamiltonian
\begin{equation}\label{eq:HM}
\mathcal{H}_M=M(x)\sigma_x,
\end{equation}
where $M(x)=M\theta(x)\theta(L-x)$. 

In the limit of a short junction, $L\ll\xi$, where $\xi=v/\Delta$ is the superconducting coherence length, and a large magnetic field, $M\gg\Delta$, the barrier is characterized by an energy-independent scattering matrix
\begin{equation}
\label{eq:Se-eq}
S_e=\left(\begin{array}{lr}r & d \\d & r\end{array}\right),
\end{equation}
where $r=-i\tanh(ML/v)$ and $d=1/\cosh(ML/v)$~\cite{Fu2009}.
Thus, the transmission probability, $D=|d|^2$, may be tuned between 0 and 1 with the height of the magnetic barrier. 
Note that the results outlined below do not actually rely on the specific form of $S_e$ in Eq.~\eqref{eq:Se-eq}, but only of the fact that it is unitary and symmetric.

Combining Eqs.~(\ref{eq:H0}-\ref{eq:HM}), the total Hamiltonian thus reads
\begin{equation}\label{eq:Hamiltonien_brut_stis_BdG}
\mathcal{H}_0=vp_x\sigma_z\tau_z+\Delta(x)e^{i\phi(x)\tau_z}\tau_x+M(x)\sigma_x.
\end{equation}

The Andreev bound states in the junction can be found by considering the following scattering problem. The wave function on either side of the junction is a superposition of right- and left-moving electrons and holes, $\Phi=(u_+,v_+,u_-,v_-)^T$, where $u$ and $v$ describe the electron and hole components in Nambu space whereas the subscripts $\pm$ refer to right- and left-movers, corresponding to up- and down-spins. Right-(Left-)moving electrons are coupled with left-(right-)moving holes via Andreev reflections. Thus, the wave function associated with a bound state at energy $\epsilon$ can be written in the form
\begin{equation}
\label{eq:ABS_WF_full}
\Phi_A(x)=\left(\begin{array}{c}
ae^{-i\varphi/2}A_A \\
A_A \\
B_A \\
ae^{i\varphi/2}B_A
\end{array}\right)e^{\kappa x},
\quad\mathrm{at}\quad x<0,  
\quad\mathrm{and}\quad
\Phi_A(x)=\left(\begin{array}{c}
C_A \\
ae^{-i\varphi/2}C_A \\
ae^{i\varphi/2}D_A\\
D_A
\end{array}\right)e^{\kappa (L-x)},
\quad\mathrm{at}\quad x>L,  
\end{equation}
where $\kappa=\sqrt{\Delta^2-\epsilon^2}/v$ and
\begin{equation}
\label{eq:coefa}
a(\epsilon)=\epsilon/\Delta-i\sqrt{1-\epsilon^2/\Delta^2}
\qquad
\mathrm{at}
\qquad
|\epsilon|<\Delta.
\end{equation}
Furthermore, right-moving electrons (holes) are coupled with left-moving electrons (holes) via spin-flip scattering at the magnetic barrier. 
Thus,
\begin{equation}
\label{eq:eigenproblem}
\left(\begin{array}{c}
B_A \\
C_A
\end{array}\right)
=
S_e
\left(\begin{array}{c}
ae^{-i\varphi/2}A_A\\
ae^{i\varphi/2}D_A \\
\end{array}\right),
\qquad
\left(\begin{array}{c}
A_A \\
D_A
\end{array}\right)
=
S_h
\left(\begin{array}{c}
ae^{i\varphi/2}B_A\\
ae^{-i\varphi/2}C_A \\
\end{array}\right),
\end{equation}
where the scattering matrix for holes is related to the scattering matrix for electrons through $S_h=-\sigma_yS_e^*\sigma_y$. The eigenproblem defined by Eqs.~\eqref{eq:eigenproblem} then defines the Andreev bound state energy and wave function.

\begin{figure}[ht!]
\begin{center}
\includegraphics[scale=0.85]{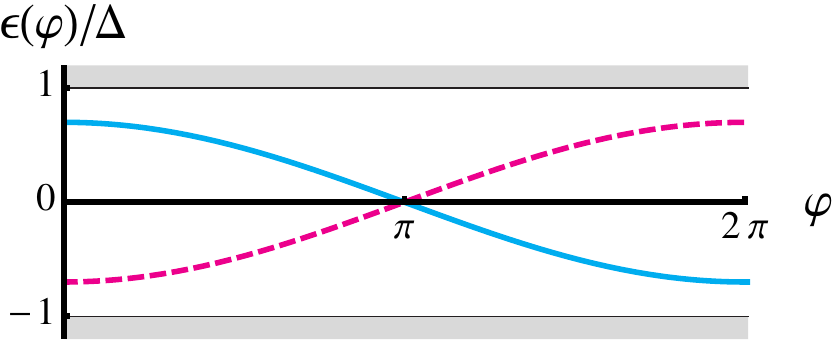}
\caption{Dependence of the energy spectrum on the superconducting phase difference $\varphi$, including the continuum of states above the gap (gray) and a filled (solid line) or empty (dashed line) Andreev bound state, in an topological Josephson junction with transparency $D=0.7$. Note that there is a ``true'' crossing at $\varphi=\pi$.}
\label{fig:Fig-spectrum}
\end{center}
\end{figure}

As a result, we find that the junction hosts a single Andreev bound state with energy
\begin{equation}\label{eq:ABS_spectrum}
\epsilon_A(\varphi)=\sqrt D\Delta \cos\frac\varphi2.
\end{equation}
The energy spectrum of the junction is shown in Fig.~\ref{fig:Fig-spectrum}. For the bound state wave function, we obtain
\begin{equation}\label{eq:WF_ABS}
B_A=D_A=-
e^{-i\varphi/2}
\left(\frac{\sqrt{1-D\cos^2\frac\varphi2}+\sqrt D\sin\frac\varphi2}{\sqrt{1-D\cos^2\frac\varphi2}-\sqrt D\sin\frac\varphi2}
\right)^{1/2}
A_A,
\end{equation}
while $C_A=A_A$. Finally, in the limit $L\to 0$ at fixed transmission $D$, the normalization condition $\int dx\;\Phi_A^\dagger\Phi_A =1$
 yields
\begin{equation}\label{eq:WF_ABS2}
|A_A|^2=
\frac\Delta{4v}
\left(\sqrt{1-D\cos^2\frac\varphi2}-\sqrt D\sin\frac\varphi2\right).
\end{equation}
We note that the bound state has equal weight on either side of the barrier, no matter what the barrier height. In particular, this remains true in the limit $D\to0$, when the two sides of the junction decouple: the single fermionic bound state is split into two Majorana fermions at zero energy, one on either side of the junction. At finite transmission, the two Majorana fermions couple and form a $4\pi$-periodic Andreev bound state, see Eq.~\eqref{eq:ABS_spectrum}. As the energy of the bound state changes sign at $\varphi=(2n+1)\pi$, where $n\in\mathbb{Z}$, the parity of the ground state of the system changes between even and odd.

The phase-dependent part of the junction energy, $E_A(\varphi)=-(n_A-1/2)\epsilon_A(\varphi)$, depends on the occupation of the bound state, $n_A$. The Josephson current carried by the bound state is given as
\begin{equation}
\label{eq:JJrelation}
I_A(\varphi)=2e \frac\partial{\partial\varphi}E_A(\varphi)=(2n_A-1)I_J\sin\frac\varphi2,
\end{equation}
where $I_J=e\sqrt D\Delta/2$. 

At fixed fermion parity $n_A$, the Josephson current \eqref{eq:JJrelation} is proportional to $\sqrt{D}$ and $4\pi$-periodic. By contrast, in conventional tunnel junctions, the Josephson current is proportional to ${D}$ and $2\pi$-periodic. This signals that, in a single-channel topological Josephson junction, the supercurrent is carried by single electrons rather than Cooper pairs, as in a conventional Josephson junction~\cite{Kwon2004}.

In equilibrium, inelastic processes that violate the conservation of the fermion parity are unavoidable. As a result, the bound state will be thermally occupied. The Josephson current in the junction is thus given as
\begin{equation}
\langle I(\varphi)\rangle=I_J\sin\frac\varphi2\tanh\left(\frac{\sqrt D\Delta}{2T}\cos\frac\varphi2\right).
\end{equation}
 In particular, at $T\ll |\epsilon_A(\varphi)|$, the system will relax to the ground state whose parity depends on the phase $\varphi$, and $\langle I(\varphi)\rangle_{T=0}=I_J|\sin(\varphi/2)|$.
 
 Note that $\langle I(\varphi)\rangle$ is $2\pi$-periodic.
The telegraph noise associated with the switching of the bound state occupation, which occurs on long time scales, results in a noisy supercurrent~\cite{Fu2009}, like in conventional junctions~\cite{noise-sns1,noise-sns2}. 
The time scale for quasiparticles above the superconducting gap to tunnel into the bound state, in the presence of a bosonic bath, was recently estimated to lie in the $\mu$s range~\cite{RainisLoss2012}, in the context of the experiment reported in~\cite{Mourik2012}.

By contrast, if the parity were fixed, the current would be $4\pi$-periodic. This was predicted to happen in a biased topological Josephson junction in the absence of inelastic processes~\cite{Kwon2004,Fu2009}. Namely, using the Josephson relation $\dot\varphi=2eV_{\rm dc}$, the junction should display a fractional Josephson effect, 
\begin{equation}
I_{\rm ac}(t)=I_J\sin\left(\frac{\omega_J}2t+\frac{\phi_0}2\right),
\end{equation}
where $\omega_J=2eV_{\rm dc}$ is the \lq\lq conventional\rq\rq\ Josephson frequency and $\phi_0$ is the phase difference at $t=0$.

By the same token, under an additional ac bias with frequency $\Omega$, the junction would display Shapiro steps when the fractional Josephson frequency $\omega_J/2=eV_{\rm dc}$ matches multiples of the applied frequency $\Omega$, namely $eV_{\rm dc}=k\Omega$ or $\omega_J=2k\Omega$ ($k\in\mathbb{Z}$). By comparison with conventional Josephson junctions, where Shapiro steps appear at $\omega_J=2eV_{\rm dc}=k\Omega$, this corresponds to the presence of the even steps only. This \lq\lq even-odd\rq\rq\ effect would be a clear signature of the $4\pi$-periodicity of the bound state~\cite{Kwon2004,Liang2011}.

However, these considerations neglect non-adiabatic processes due to the applied bias. Namely, the applied bias leads to a dynamic coupling between the bound state and the continuum of states outside the gap. There are two different ways to approach this problem. Starting from the bound state spectrum, one may consider the probability to change the occupation of the bound state due to non-adiabatic transitions between the bound state and the continuum. Alternatively, on may abandon the image of a bound state altogether and consider scattering states due to multiple Andreev reflections. We will discuss both approaches in the following chapters.

While we concentrate on the specific model of the junction introduced above, the main conclusions are more general. Note that the model for a topological Josephson junction based on a nanowire with strong spin-orbit coupling $B_{\rm SO}$ in the presence of a Zeeman field $B_Z$~\cite{Lutchyn2010,Oreg2010} is more complex, but reduces to the above model in the limit $B_Z\gg\Delta,\mu,B_{\rm SO}$. 
Moreover, as time-reversal symmetry is broken, a non-magnetic barrier is sufficient to induce backscattering. Thus, the height of the barrier in a nanowire-based topological junction may be controlled with an electrostatic gate. Further differences arise when taking into account the finite length of the wire and/or the presence of multiple channels. We will comment on these effects in the next section.
 
\section{Bound state dynamics}\label{sec:section_2}

In this section, we consider a phenomenological model of the bound state dynamics. Similar models have been used in {\cite{Nazarov-noise,Aguado-transients,Dominguez2012,Sau2012,virtanen-recher}}. Here the bound state dynamics is due to a non-adiabatic coupling with the continuum. We define a characteristic switching time $\tau_{\rm s}$ over which the occupation of the bound state changes. This time scale has then to be compared with the characteristic time scale $\tau_{\mathcal R}$ set by the external circuit over which the phase difference across the Josephson junction may adjust. The ratio between these two time scales determines the visibility of experimental signatures of the fractional Josephson effect in the Shapiro steps and the finite-frequency current noise.

\subsection{Phenomenology of the bound state dynamics}\label{subsec:subsection_21} 

We will consider a junction with sufficiently high transparency such that the minimal distance in energy between the bound state and the continuum, $\delta=\Delta(1-\sqrt D)$, is much smaller than the gap $\Delta$. In that case, one may assume that the coupling between the bound state and the continuum occurs in narrow intervals of $\varphi$ around $2n\pi$.\footnote{We will show later in Sec.~\ref{sec5} that this is indeed the case.} Thus, the occupation probability $P_n$ of the bound state is fixed at phases $\varphi_n<\varphi<\varphi_{n+1}$, where $n=\rm{Int}\left[\varphi/2\pi\right]$.

If the bound state is filled, the particle may escape to the continuum with a probability $s$ when the bound state  approaches the empty states above the gap at $\varphi_{2n}=4n\pi$. If the bound state is empty, there is a probability $s$ for a particle from the continuum to occupy it when the bound state approaches the filled states below the gap at phases  $\varphi_{2n+1}=(4n+2)\pi$. We, thus, may write the following equations linking the probabilities $P_n$ and $Q_n=1-P_n$ in neighboring phase intervals:
\begin{subequations}\label{eq:conditional_probabilities}
\begin{equation}
\left(\begin{array}{c}P_{2n}\\Q_{2n}\end{array}\right)=\left(\begin{array}{cc}1-s&0\\s&1\end{array}\right)\left(\begin{array}{c}P_{2n-1}\\Q_{2n-1}\end{array}\right)
\end{equation}
and
\begin{equation}
\left(\begin{array}{c}P_{2n+1}\\Q_{2n+1}\end{array}\right)=\left(\begin{array}{cc}1&s\\0&1-s\end{array}\right)\left(\begin{array}{c}P_{2n}\\Q_{2n}\end{array}\right).
\end{equation}
\label{eq-proba}
\end{subequations}
Under dc bias, the phase increases with time as $\varphi(t)=2eV_{\rm dc}t+\phi_0$. To find the probability of the state being occupied at times $(\varphi_{n+k}-\phi_0)/(2eV_{\rm dc})<t<(\varphi_{n+k+1}-\phi_0)/(2eV_{\rm dc})$, we solve equations \eqref{eq-proba} iteratively to obtain
\begin{equation}
P_{n+k}=P_{n+k}^{\infty}+\left(1-s\right)^{k}\left(P_n-P_n^{\infty}\right),
\end{equation}
where $\varphi_n<\phi_0<\varphi_{n+1}$.

At $k\gg-1/\ln\left(1-s\right)$, corresponding to times $t\gg\tau_{\rm s}=-2\pi/\left[eV_{\rm{dc}}\ln\left(1-s\right)\right]$, the occupation probability approaches the long-time value $P_{n+k}^{\infty}=\left[1-(-1)^{n+k}s/\left(1-s\right)\right]/2$. Note that $P_{n+k}^\infty$ is $4\pi$-periodic and independent of the initial state, reflecting the Markovian property of the time evolution.

Thus, we have identified the characteristic time scale $\tau_{\rm s}$ over which the occupation of the bound state switches. To understand the effect of this switching on measurable quantities, we have to compare this time scale with other relevant time scales  of the system. It turns out that the most important time scale is the phase adjustment time $\tau_{\cal R}$, set by the circuit the Josephson junction is embedded in. In order to identify this time scale, we use an RSJ model.

\subsection{RSJ-Model}

\begin{figure}[ht!]
\begin{center}
\includegraphics[scale=0.3]{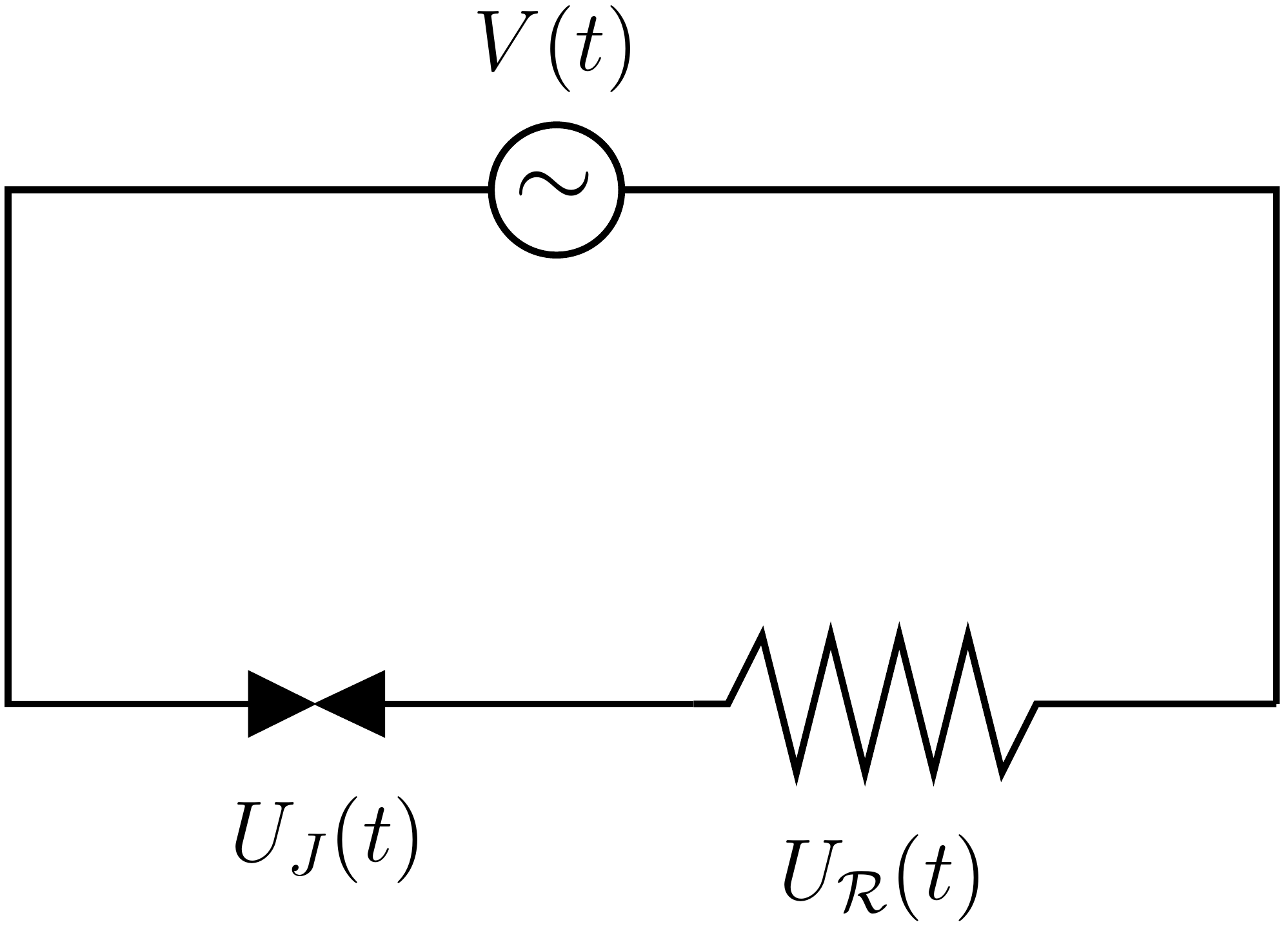}
\caption{Electrical circuit consisting of a voltage-biased Josephson junction in series with an external resistance. Here $V(t)$ is the bias voltage, $U_J(t)=\dot\varphi(t)/(2e)$ is the voltage at the junction, and $U_\mathcal{R}(t)={\cal R} I_S(t)$ is the voltage at the resistance.}
\label{fig:circuit}
\end{center}
\end{figure}

While RSJ stands for \lq\lq resistively-shunted Josephson junction\rq\rq, the same model also applies to a voltage-biased Josephson junction in series with an external resistance ${\cal R}$, cf.~Fig.~\ref{fig:circuit}. In that case,
\begin{equation}
V(t)={\cal R}I_S(t)+\frac1{2e}\dot\varphi(t),
\label{eq-RSJ}
\end{equation}
where $V(t)$ is the applied bias and $I_S(t)$ is the Josephson current. For our topological Josephson junction, $I_S(t)=(-1)^{n_A(t)} I_J\sin\left(\varphi(t)/2\right)$, where $n_A(t)=0$ or $1$ is the occupation of the bound state.

In order to study Shapiro steps, we will consider combined dc and ac voltages, $V(t)=V_{\rm dc}+V_{\rm ac}\cos(\Omega t)$. For dc voltages close to multiples of the microwave frequency, $eV_{\rm dc}\sim k\Omega$ with $k\in\mathbb{Z}$, the phase may be decomposed into a rapidly varying part and a slowly varying part {$\chi(t)$} that adjusts to the external circuit. Namely,
\begin{equation}
\varphi(t)=2k\Omega t +\frac{2eV_{\rm ac}}\Omega\sin\left(\Omega t\right)+\chi(t).
\end{equation}
Substituting this decomposition into Eq. \eqref{eq-RSJ}, we find
\begin{equation}
eV_{\rm dc}-k\Omega\simeq(-1)^{n_A(t)} e{\cal R}I_JJ _{-k}\left(\frac{eV_{\rm ac}}\Omega\right)\sin\frac{\chi(t)}2+\frac1{2}\dot\chi(t).
\label{eq-tauR}
\end{equation}
From Eq. \eqref{eq-tauR} we can extract the characteristic time scale~\cite{Sau2012} for the evolution of $\chi$. Namely, $\tau_{\cal R}^{(k)}=1/|e{\cal R}I_JJ _{-k}(\alpha)|$ with $\alpha=eV_{\rm ac}/\Omega$.

If $\tau_{\rm s}\gg\tau_{\cal R}$, the occupation of the bound state remains constant over the typical time scale over which the phase $\chi(t)$ adjusts. Thus, in this limit, we may solve Eq.~\eqref{eq-tauR} at fixed $n_A(t)$. On the other hand, if $\tau_{\rm s}\ll\tau_{\cal R}$, the Josephson current switches randomly before the phase $\chi(t)$ may adjust. Below we will discuss the consequences for the average current, in particular the Shapiro steps, as well as the finite-frequency noise.

\subsection{Even-odd effect in Shapiro steps}\label{subsec:subsection_22}

In a Josephson junction under dc and ac bias, Shapiro steps may appear at discrete values of the voltage $V_{\rm dc}$~\cite{Shapiro1963}. In order to be able to use our phenomenological model, introduced in Sec.~\ref{subsec:subsection_21}, we need to make the following assumptions:
\begin{itemize}
\item $V_{\rm ac}\ll V_{\rm dc}$ such that the phase velocity $\dot\varphi=2eV(t)\approx 2eV_{\rm dc}$. This condition ensures that the switching parameter $s$ is not significantly modified by the additional ac bias.
\item $\Omega\ll\delta$ such that multi-photon processes are required to excite particles between the bound state and the continuum. This condition ensures that the coupling between the bound state and the continuum is dominated by the non-adiabatic processes due to the finite phase velocity considered above.
\end{itemize}
In the limit of short phase adjustment time, $\tau_{\mathcal R}\ll\tau_{\rm s}$, the switching can be neglected and the current can be obtained from Eq.~\eqref{eq-tauR} with $n_A(t)=n_0$ fixed. For voltages sufficiently close to $k\Omega$, namely $|eV_{\rm dc}-k\Omega|<1/\tau_{\cal R}^{(k)}$, Eq.~\eqref{eq-tauR} admits the constant solution
\begin{equation}
\chi(t)=\bar\chi\equiv 2(-1)^{n_0}\arcsin\left((eV_{\rm dc}-k\Omega)\tau_{\cal R}^{(k)}\right),
\end{equation}
corresponding to a current $I_{\rm dc}^<=\frac1{e{\cal R}}\left\{eV_{\rm dc}-k\Omega\right\}$.

For $|eV_{\rm dc}-k\Omega|>1/\tau_{\cal R}^{(k)}$, the dc component of the current may be obtained by integrating Eq.~\eqref{eq-tauR} over one period. One finds
\begin{equation}
\label{eq-T}
T= \frac{2\pi\tau_{\cal R}^{(k)}}{\sqrt{\left[(eV_{\rm dc}-k\Omega)\tau_{\cal R}^{(k)}\right]^2-1}},
\end{equation}
whereas the dc current is given as
\begin{equation}
I_{\rm dc}^>\simeq (-1)^{n_0}I_JJ _{-k}\left(\alpha\right)\overline{\sin \frac{\chi(t)}2}
=\frac1{e{\cal R}}\left\{eV_{\rm dc}-k\Omega-\frac12\overline{\dot\chi(t)}\right\},
\label{eq-Idcg}
\end{equation}
where the bar denotes time averaging.  With $\overline{\dot\chi(t)}=4\pi/T$, 
 Eqs.~\eqref{eq-T} and \eqref{eq-Idcg} yield the current $I_{\rm dc}^>=\frac1{e{\cal R}}\Big\{eV_{\rm dc}-k\Omega$ $-\sqrt{(eV_{\rm dc}-k\Omega)^2-(\tau_{\cal R}^{(k)})^{-2}}\Big\}$.

Combining the different regimes, we finally obtain~\cite{S-quantronics}
\begin{equation}
\label{eq-current-RSJ}
I_{\rm dc}=\sum_k \frac{\delta V_k}{{\cal R}}\left\{1-\theta\left[1-\left(\frac{{\cal R}I_k}{\delta V_k}\right)^2\right]\sqrt{1-\left(\frac{{\cal R}I_k}{\delta V_k}\right)^2}\right\},
\end{equation}
where $I_k=I_J|J_{k}(\alpha)|$ is the height of the Shapiro step at $eV_{\rm dc}=k\Omega$ and $\delta V_k=V_{\rm dc}-k\Omega/e$. 

Eq. (\ref{eq-current-RSJ}) shows the expected \lq\lq even-odd\rq\rq\ effect~\cite{Kwon2004,Liang2011,Dominguez2012} (see discussion in Sec.~\ref{sec:section_1}), namely Shapiro steps appear at voltages $eV_{\rm dc}=k\Omega=2k\times(\Omega/2)$ only. 

Let us now consider the opposite limit $\tau_{\rm{s}}\ll\tau_{\cal R}$. In that case, the occupation switches much faster than the phase across the junction can adjust. We may, thus, neglect the phase adjustment due to the external circuit and compute the current using the long-time probabilities $P^{\infty}/Q^{\infty}$ obtained in Sec.~\ref{subsec:subsection_21}. The average current at times $t\gg\tau_{\rm{s}}$ then reads
\begin{equation}\label{eq:courant_long_times}
\left\langle I(t)\right\rangle=I_J\sin{\frac{\varphi(t)}{2}}\left[Q_{{\rm Int}\left[\varphi(t)/2\pi\right]}^{\infty}-P_{{\rm Int}\left[\varphi(t)/2\pi\right]}^{\infty}\right]=\frac{sI_J}{2-s}\left|\sin{\frac{\varphi(t)}{2}}\right|.
\end{equation}
There are two important things to note about this result. {\em (i)} The average current is $2\pi$-periodic. Due to the random switching of the bound state occupation, the $4\pi$-periodicity associated with the conservation of parity has disappeared. {\em (ii)}  The average current is strongly suppressed for small switching probabilities $s\ll1$, namely $\left\langle I(t)\right\rangle\propto s$. 

Extracting the dc component of the current from Eq.~\eqref{eq:courant_long_times} under applied dc and ac bias shows that Shapiro steps are strongly suppressed. Furthermore, the $2\pi$-periodicity of Eq.~\eqref{eq:courant_long_times} implies that the \lq\lq even-odd\rq\rq\ effect is absent.\footnote{Note, however, that Eq.~\eqref{eq:courant_long_times} does not allow us to obtain the exact shape of Shapiro steps in this regime. Namely the result for the dc component of the current obtained from Eq.~\eqref{eq:courant_long_times} depends on the initial phase. In a realistic circuit, this dependence would disappear at times $t\gg\tau_{\cal R}$.}
Observing the fractional Josephson effect via Shapiro step measurements thus requires $\tau_{\rm{s}}\gg\tau_{\cal R}$. 

As both $\tau_{\rm{s}}$ and $\tau_{\cal R}$ are voltage-dependent, this condition may differ for different Shapiro steps. A recent experiment reported the suppression of the first Shapiro step in a nanowire-based Josephson junction at large magnetic field~\cite{Rokhinson2012}. This was interpreted as a manifestation of the ``even-odd'' effect, signaling Majorana bound states in the junction. The observation that the third step, however, was not suppressed could be consistent with the decrease of $\tau_{\rm{s}}$ with increasing $V_{\rm dc}$, under the assumption that on the first step $\tau_{\rm{s}}$ is larger than the phase adjustment time, whereas on the third step the situation is reversed. However, the experimentally studied junction had many channels and, thus, contained a large $2\pi$-periodic harmonic, in contrast with our single-channel model. Therefore our results cannot be directly applied to the experiment~\cite{Rokhinson2012}. A more quantitive theoretical description of that experiment may be found in \cite{Dominguez2012}.

\subsection{Current noise}\label{subsec:subsection_23}

As discussed in the previous section, signatures of the $4\pi$-periodic bound state are absent in the average current, if the switching time is faster than the phase adjustment time. We thus turn to current fluctuations in this regime. Namely, while the average current is sensitive only to the long-time properties, the finite-frequency noise allows one to probe correlations at shorter times.

In particular, we consider the current noise spectrum,
\begin{equation}\label{eq:current_current_correlation}
S(\omega)=\int{d\tau\,e^{i\omega\tau}\overline{\left\langle\delta I(t)\delta I(t+\tau)+\delta I(t+\tau)\delta I(t)\right\rangle}},
\end{equation}
where $\delta I(t)=I(t)-\left\langle I(t)\right\rangle$ is the deviation from the statistical average. The noise may be obtained via the correlator 
\begin{equation}
\left\langle I(\varphi_1)I(\varphi_2)\right\rangle=I_J^2\sin{\frac{\varphi_1}2}\sin{\frac{\varphi_2}2}\left[Q_{n_1}^{\infty}x_{n_2}(P_{n_1}=0)-P_{n_1}^{\infty}x_{n_2}(P_{n_1}=1)\right],
\end{equation}
at $\varphi_1<\varphi_2$, where $n_i={\rm{Int}}\left[\varphi_i/(2\pi)\right]$ and $x_{n_i}=Q_{n_i}-P_{n_i}$. Using the conditional probabilities obtained from Eq.(\ref{eq:conditional_probabilities}), the  correlator evaluates to 
\begin{equation}
\left\langle I(\varphi_1)I(\varphi_2)\right\rangle=4I_J^2\frac{1-s}{\left(2-s\right)^2}\sin{\frac{\varphi_1}{2}}\sin{\frac{\varphi_2}{2}}\left(1-s\right)^{n_2-n_1}.
\end{equation}
Let us consider the dc case first. Using $\varphi_i=2eV_{\rm dc}t_i+\phi_0$,  Eq. (\ref{eq:current_current_correlation}) yields 
\begin{equation}
\label{eq:Svsomega}
S(\omega)=\frac{4sI_J^2}{\pi(2-s)} \frac{(eV_{\rm dc})^3}{[\omega^2\!-\!(eV_{\rm dc})^2]^2}\frac{4\cos^2\!\frac{\pi \omega}{2eV_{\rm dc}}}{4\cos^2\!\frac{\pi\omega}{2eV_{\rm dc}}+\frac{s^2}{1- s}}.
\end{equation}
If $s\ll 1$, Eq.~\eqref{eq:Svsomega} simplifies to
 \begin{equation}
S(\omega)\simeq \frac{I_J^2}2\frac{seV_{\rm dc}/\pi}{(\omega\mp eV_{\rm dc})^2+(seV_{\rm dc}/\pi)^2}
\label{eq-fin_simp}
\end{equation}
at $|\omega\mp eV_{\rm{dc}}|\ll eV_{\rm{dc}}$. Eq.~\eqref{eq-fin_simp} shows that the noise spectrum has sharp peaks at $\omega=\pm eV_{\rm{dc}}$, i.e., at half of the \lq\lq usual\rq\rq\ Josephson frequency. The position of the peak reveals the $4\pi$-periodicity of the Andreev bound state. Namely, the noise is sensitive to the transient $4\pi$-periodic behavior \cite{Aguado-transients} of the current at times smaller than the lifetime of the bound state. Between two switching events, the current oscillates with the fractional Josephson frequency. Therefore, the noise spectrum which probes the current correlations at different times shows a peak whose inverse width is proportional to the survival time $\tau_{\rm{s}}$ of the fractional Josephson effect. Thus, not only does the noise probe the $4\pi$-periodicity of the bound state, but it also allows one to estimate the lifetime of that bound state.

Note that the Markovian model developed above is also applicable to nanowires with strong spin-orbit coupling and a Zeeman energy much larger than the superconducting gap. In finite length wires, the presence of additional Majorana modes at the ends of the wire splits the zero-energy crossing at  $\varphi=(2n+1)\pi$. Thus, in addition to the non-adiabatic processes that we considered, non-adiabatic processes in the vicinity of the avoided crossing become important~\cite{Nazarov-noise,Aguado-transients,Dominguez2012}. In that case, the underlying $4\pi$-periodicity would be visible only if the probability of Landau-Zener tunneling across the gap at $\varphi=(2n+1)\pi$ is large while the switching probability due to the coupling with the continuum at $\varphi=2n\pi$ remains small.\footnote{The same physics also applies to non-topological junctions if the gap between the bound states at $\varphi = (2n + 1)\pi$ is much smaller than the gap to the continuum at $\varphi = 2n \pi$ \cite{Sau2012,exp-4pi}.}

In conclusion, measuring the finite frequency current noise could be a direct experimental evidence of the fractional Josephson effect and, thus, of the presence of Majorana fermions in the junction. Even if the current noise spectrum at high frequencies can be hard to obtain experimentally {\cite{freq-noise,Deblock2003}, the addition of a small ac voltage with frequency $\Omega$ may shift the peak to lower frequencies, namely $\omega=\pm(eV_{\rm dc}-k\Omega)$. 

\section{Current in terms of Multiple Andreev reflections}\label{sec:section_3}

In section \ref{sec:section_2}, we investigated the dynamics of the Andreev bound state in a topological Josephson junction at low voltages and high transparencies of the junction. We showed that the most robust signature of the $4\pi$-periodicity of the bound state is a peak in the finite-frequency current noise at $\omega=eV_{\rm dc}$. In this section, we take a different approach and investigate the I-V characteristics of a voltage-biased topological Josephson junction using the scattering formalism. This allows us to study the peak in the noise spectrum at arbitrary voltages and transparencies.

\subsection{Scattering matrix approach}\label{subsec:subsection_31}

Our starting point is the topological Josephson junction described by the Hamiltonian (\ref{eq:Hamiltonien_brut_stis_BdG}). To incorporate the finite bias, we have to add the following term to the Hamiltonian,
\begin{equation}
{\cal H}_U=-eU(x,t)\tau_z,
\label{eq-HU}
\end{equation}
where $U(x,t)=V(t)\left[\theta(-x)-\theta(x-L)\right]/2$. In the following, we will restrict our attention to a dc bias $V(t)=V_{\rm dc}$. Using the Josephson relation, the phase $\varphi(t)$ in Eq.~\eqref{eq:Hamiltonien_brut_stis_BdG} is then given as $\varphi(t)=2eV_{\rm dc}t+\phi_0$. For simplicity, we will set $\phi_0=0$.

To investigate the transport properties of the junction, we adopt the Landauer-B\"uttiker formalism of coherent quantum transport, also used to address the transport properties of conventional Josephson junctions \cite{Averin1995,Bratus1995,Cuevas1996}.
In this approach, the superconducting electrodes are quasiparticle reservoirs in local thermodynamic equilibrium whereas the junction can be described by a scattering matrix, see section \ref{sec:section_1}.

\begin{figure}[ht!]
\begin{center}
\includegraphics[scale=0.35]{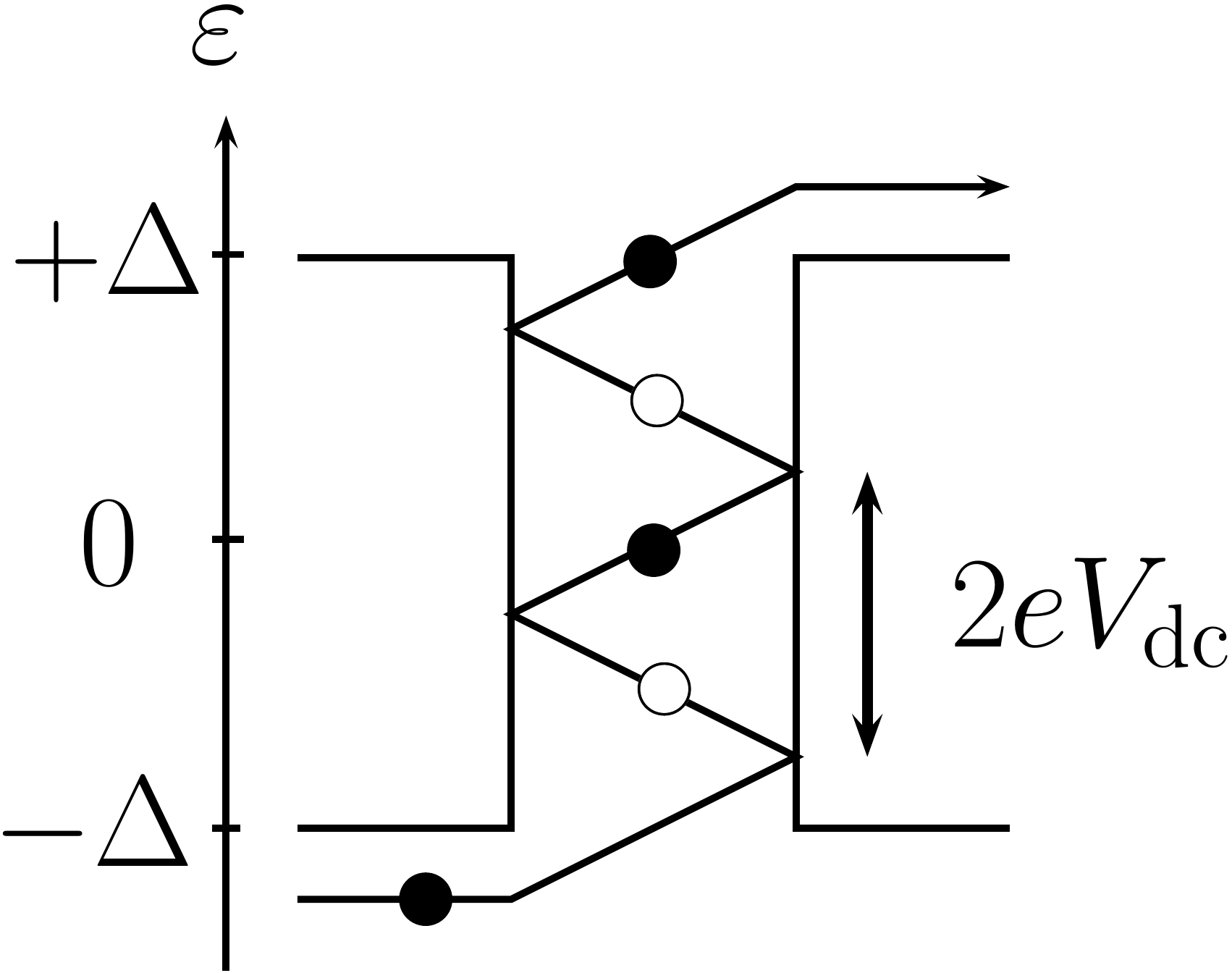}
\caption{Schematic representation of a Multiple Andreev Reflexion (MAR) process.}
\label{fig:MAR}
\end{center}
\end{figure}

Using the unitary transformation $\mathcal U(t)=\exp{\left[i\phi(t)\tau_z/2\right]}$, we work in a gauge with zero electric potential in the electrodes.\footnote{We recall that the Hamiltonian $\mathcal H_0+\mathcal H_U$ transforms into 
$\mathcal H=\mathcal U(t)^\dagger(\mathcal H_0+\mathcal H_U)\mathcal U(t)-i\mathcal U(t)^\dagger\mathcal{\dot U}(t)$ 
under the time-dependent unitary transformation 
$\mathcal U(t)$.} This allows us to transfer the time dependence of the Hamiltonian ${\cal H}_U$ to the scattering matrix \eqref{eq:Se-eq}. Namely,
\begin{subequations}
\begin{equation}
S_e(t)=\left(\begin{array}{lr}r & de^{ieV_{\rm dc}t} \\de^{-ieV_{\rm dc}t} & r\end{array}\right).
\end{equation}
The scattering matrix for holes is related to the scattering matrix for electrons via 
\begin{equation}
S_h(t)=-\sigma_yS_e^*(t)\sigma_y.
\end{equation}
\end{subequations}
The oscillating off-diagonal elements of the scattering matrix reflect the time dependence of the problem and result in an inelastic scattering of quasiparticles at each traversal of the barrier. Due to Andreev reflections, the quasiparticles may traverse the junction multiple times before being transmitted into the reservoirs, a process called multiple Andreev reflections (MAR) and illustrated in Fig.~\ref{fig:MAR}. 
As a consequence, the scattering states are a superposition of states with energies $\epsilon+2neV_{\rm dc}$ with $n\in\mathbb{Z}$. For instance, the wave function of an incoming electron ($e$) {with energy $\epsilon$} from the left ($l$) reservoir can be written in the form
\begin{equation}
\Phi_{\epsilon}^{el}(0,t)=\frac J{\sqrt{2\pi v}}\sum_n\left(\begin{array}{c}
\delta_{n0}+a_{2n}A_n \\
A_n \\
B_n \\
a_{2n} B_n
\end{array}\right)
e^{-i(\epsilon+2neV)t}, \qquad
\Phi_{\epsilon}^{el}(L,t)=\frac J{\sqrt{2\pi v}}\sum_n\left(\begin{array}{c}
C_n \\
a_{2n+1}C_n \\
a_{2n+1}D_n \\
D_n
\end{array}\right)
e^{-i[\epsilon+(2n+1)eV]t}.
\end{equation}
Here $a_n(\epsilon)=a(\epsilon+neV_{\rm dc})$ with $a(\epsilon)$ as defined in Eq.~\eqref{eq:coefa} for $|\epsilon|<\Delta$ and $a(\epsilon)=\epsilon/\Delta-\mathrm{sign}(\epsilon)\sqrt{\epsilon^2/\Delta^2-1}$ for $|\epsilon|\geq\Delta$. Furthermore, $J(\epsilon)=\sqrt{1-|a(\epsilon)|^2}$. The wave functions for holes or particles incoming from the right reservoir differ by the position of the source term $\propto \delta_{n0}$.

Rather than a single set of coefficients $A,B,C,D$ as in the equilibrium case, we now have an infinite number of coefficients $A_n,B_n,C_n,D_n$ ($n\in\mathbb{Z}$) which are related through the set of equations
\begin{equation}
\left(\begin{array}{c}
B_n \\
C_n
\end{array}\right)
=
S_e(0)
\left(\begin{array}{c}
\delta_{n,0}+a_{2n}A_n\\
a_{2n+1}D_n \\
\end{array}\right),
\qquad
\left(\begin{array}{c}
A_n \\
D_{n-1}
\end{array}\right)
=
S_h(0)
\left(\begin{array}{c}
a_{2n}B_n\\
a_{2n-1}C_{n-1} \\
\end{array}\right).
\end{equation}
As the coefficients decrease with increasing $|n|$, the set of equations  may be truncated at some $|n|=N_{\rm max}$ and then solved numerically.

To obtain the current, we express the current operator $\hat I=ev_F[\hat\psi_+^\dagger(0)\hat\psi_+(0)-\hat\psi_-^\dagger(0)\hat\psi_-(0)]$ through the scattering states. Namely, using a Bogoliubov transformation, $\hat\psi_s(x)
=
\sum_\nu
\left[
u_{s\nu}(x)
\hat\gamma_\nu
-s v^*_{-s\nu}(x)
\hat\gamma^\dagger_\nu
\right]$,
where $s=\pm$ and $\nu=\{\epsilon,i,\alpha\}$ labels an incoming state with positive energy $\epsilon$, from the lead $i=l,r$, and of the type $\alpha=e,h$. Using the scattering wave functions $\Phi_{\epsilon}^{\alpha i}=(u_{+\nu},v_{+\nu},u_{-\nu},v_{-\nu})^T$ given above, the average current in the stationary regime takes the form
\begin{equation}
I(t)=\langle\hat I(t)\rangle=\sum_n I_n e^{i2neV_{\rm dc}t},
\label{eq-I_n}
\end{equation}
where
\begin{equation}
I_n=
\frac eh
\Big\{
DeV_{\rm dc}\delta_{n0}
-\int d\epsilon\, \tanh\frac{\epsilon}{2T}
J^2
\big[a^*_{2n}A^*_n+a_{-2n}A_{-n}
+
\sum_m(1+a^*_{2(m+n)}a_{2m})
\left(A^*_{m+n}A_m-B^*_{m+n}B_m\right)
\big]\Big\}.
\label{eq:current}
\end{equation}
Similarly, the current noise \eqref{eq:current_current_correlation} can be expressed in terms of the coefficients $A_n,B_n,C_n,D_n$~\cite{Badiane2011}.

\subsection{Ac current}

Let us first consider the average ac current. As can be seen from Eq.~\eqref{eq-I_n} only the usual Josephson harmonics appear. From the earlier discussion in section \ref{sec:section_2}, this was to be expected. Namely, in the long-time limit, $t\gg\tau_{\rm s}$, random switching of the occupation of the bound state averages out the fractional Josephson effect.

\begin{figure}[ht!]
a)\includegraphics[width=0.46\textwidth]{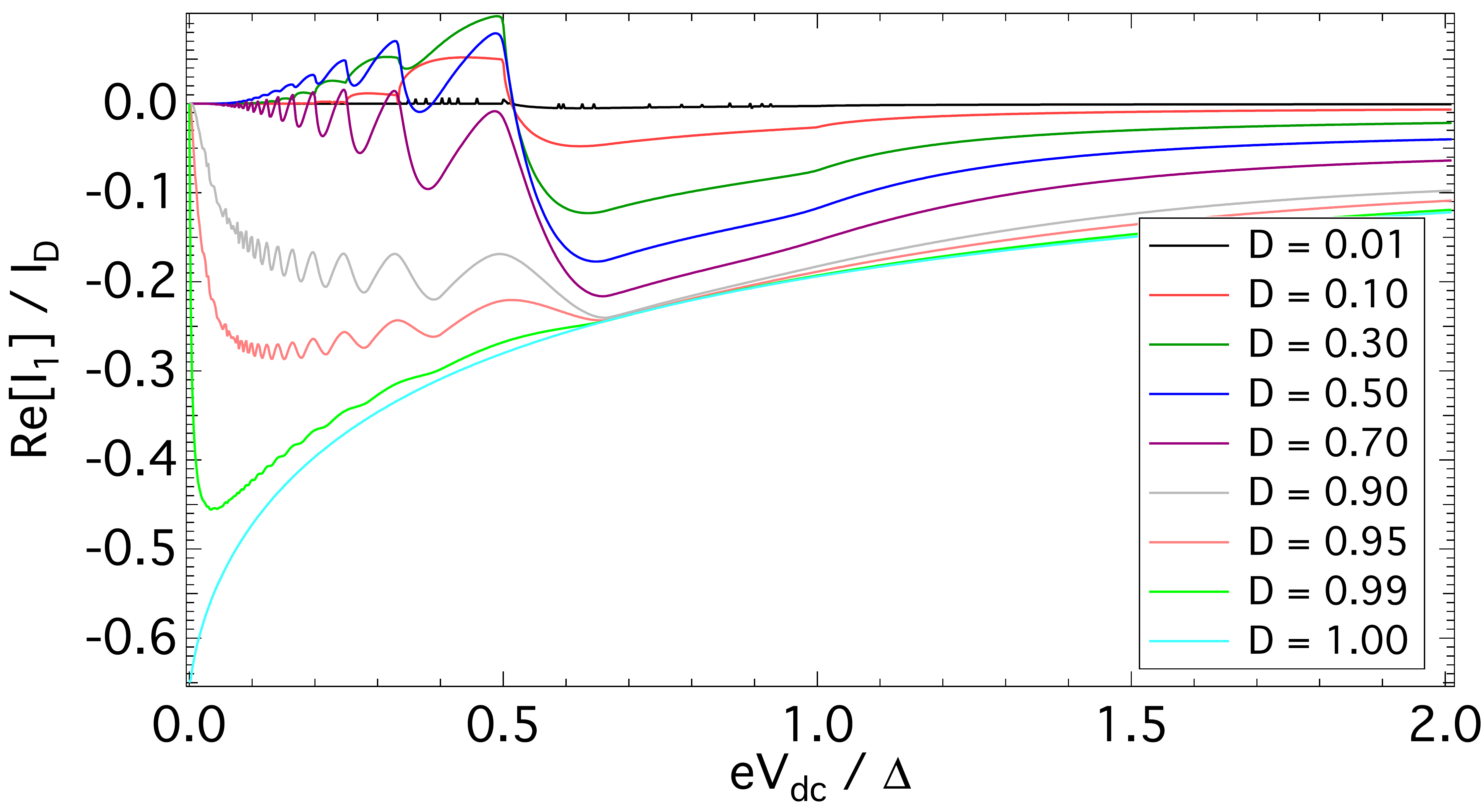}
\hfill b)\includegraphics[width=0.46\textwidth]{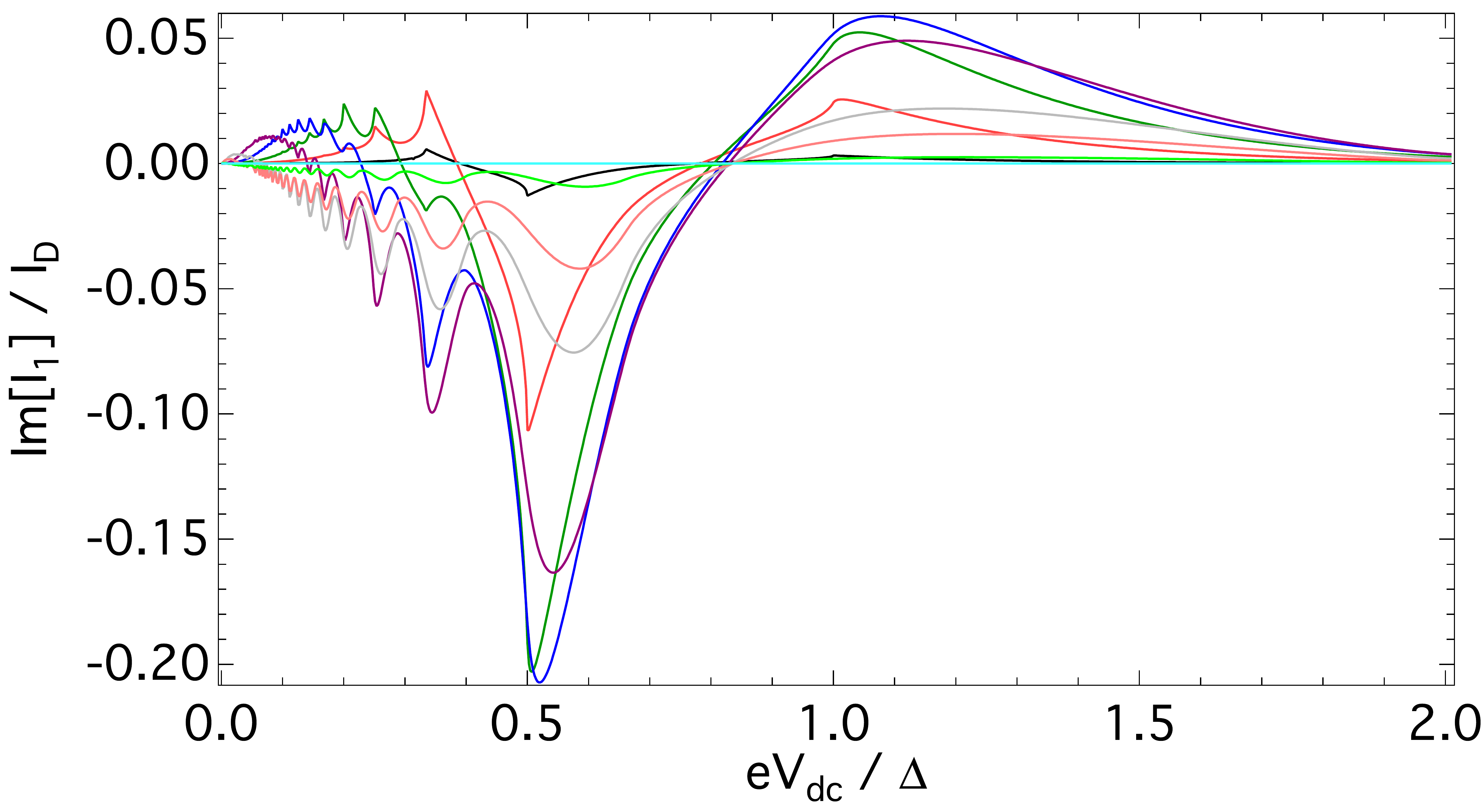}
\caption{Real and imaginary part of the first harmonic $I_1$ of the ac current at the conventional Josephson frequency $\omega_J=2eV_{\rm dc}$. Here $I_D=G_N\Delta/e$ with $G_N=De^2/h$. The legend shown in a) applies to both figures. Note that $I_1$ vanishes in the limit $V_{\rm dc}\to0$ and/or $D\to0$. 
}
\label{fig:I1}
\end{figure}

Thus, the lowest harmonic is $I_1$, oscillating at the Josephson frequency $\omega_J$. Its real and imaginary parts are shown in Fig.~\ref{fig:I1}. While the behavior at finite voltage and arbitrary transmission is more complicated, we note two main features. In the limit $V_{\rm dc}\to0$, the ac current $I_1$ vanishes for all transparencies $D<1$. This behavior is consistent with our earlier results in terms of the bound state dynamics: As the voltage approaches zero, the phase velocity becomes smaller and smaller. Thus, non-adiabatic processes become more and more suppressed. Therefore, the $2\pi$-periodic current $I_1$ vanishes. Furthermore, in the limit $D\to0$, the ac current vanishes at all voltages. As the transmission decreases, the gap between the bound state and the continuum increases. This leads to a suppression of non-adiabatic processes even at higher voltages, and therefore to a suppression of $I_1$.

\subsection{Finite-frequency current noise spectrum}\label{subsec:subsection_33}

We now turn to the current noise. As the complete formula is not very instructive, we do not show it here, but refer the reader to Ref.~\cite{Badiane2011}. The numerical results for the finite-frequency noise are shown in Fig.~\ref{fig:2fig}.

In Fig.~\ref{fig:2fig} a), corresponding to a transmission $D=0.2$, a peak at $\omega=eV_{\rm dc}$ is clearly visible for voltages up to the gap $\Delta$. In Fig.~\ref{fig:2fig} b), corresponding to a higher transmission $D=0.6$, the peak at $\omega=eV_{\rm dc}$ is distinct only for small voltages whereas it becomes very broad for larger voltages. As discussed earlier, the width of the peak can be related to the inverse of the lifetime of the bound state. The decrease of the lifetime with increasing voltage or transmission is reflected in the increasing broadening of the peak.

\begin{figure}[ht!]
a)\includegraphics[width=0.46\textwidth]{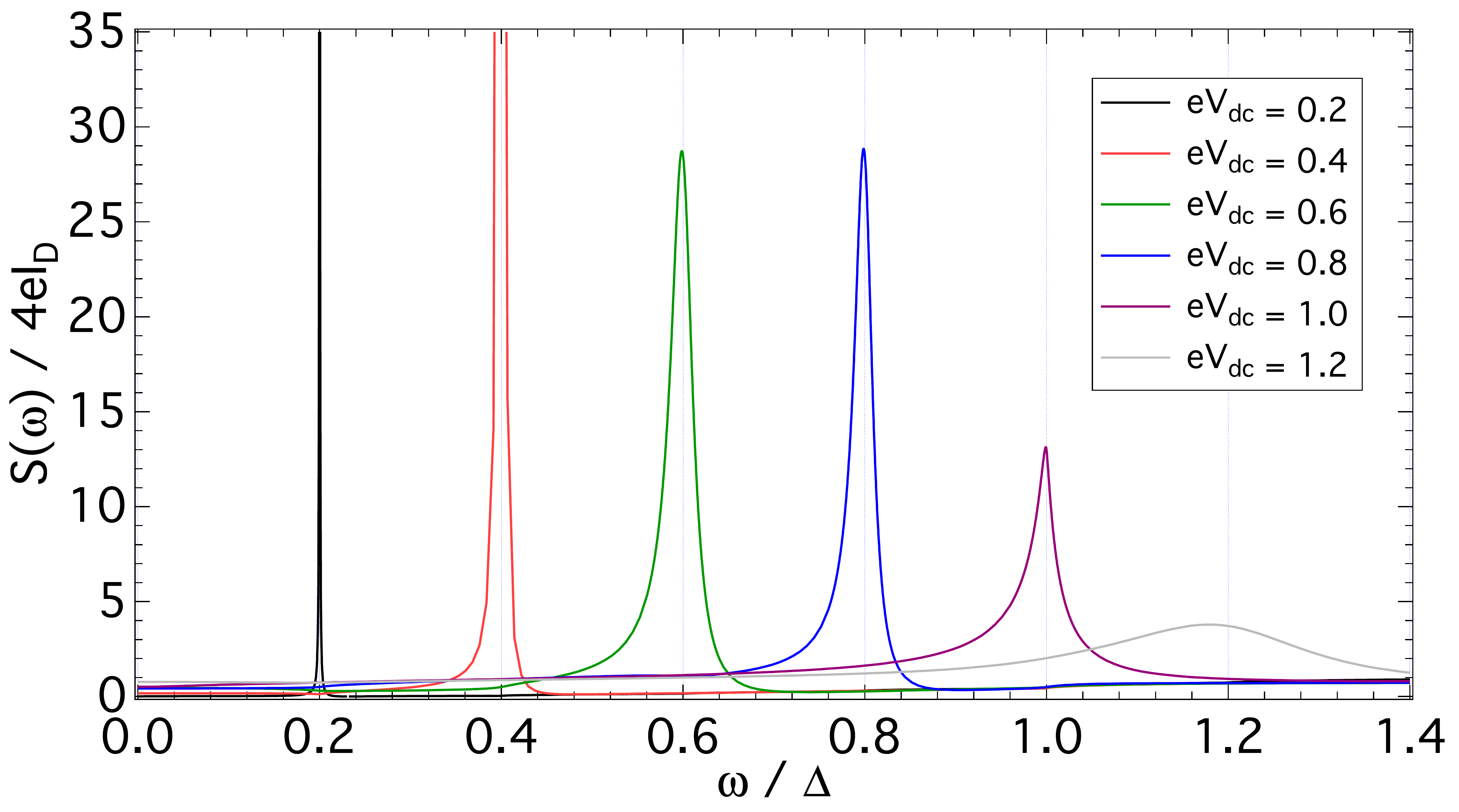}
\hfill b) \includegraphics[width=0.46\textwidth]{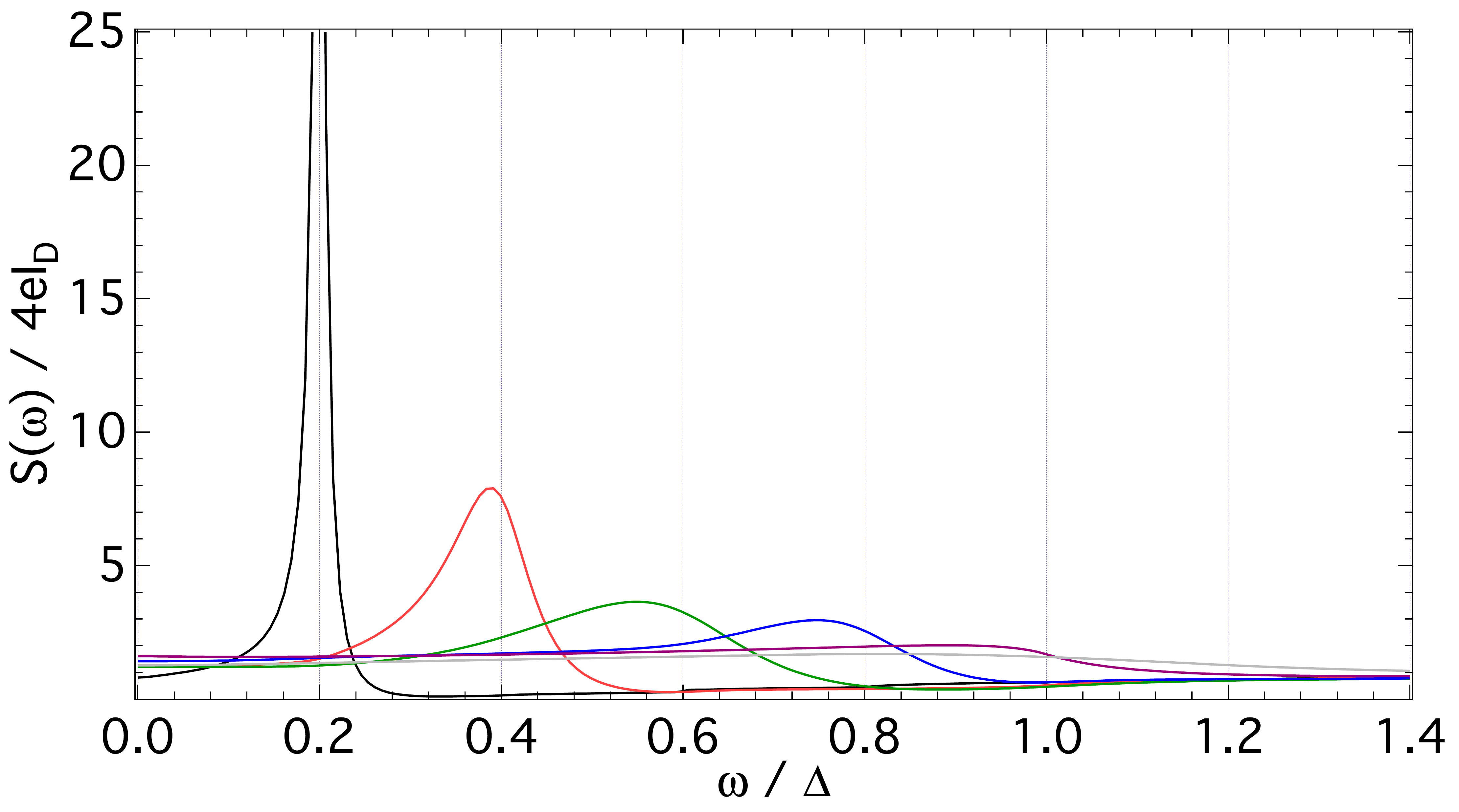}
\caption{Finite-frequency current noise $S(\omega)$ for different bias voltages with a) $D=0.2$ and b) $D=0.6$.  The legend shown in a) applies to both figures. Note the peak at $\omega=eV_{\rm dc}$, which widens with increasing voltage and/or transmission. 
}
\label{fig:2fig}
\end{figure}

These results extend the conclusions of Sec. \ref{sec:section_2} about the observability of the fractional Josephson effect via the noise spectrum to arbitrary voltages and transmissions. The peak in the noise spectrum should be visible as long as the lifetime of the bound state is much longer than the Josephson period.

\section{Switching rate in almost ballistic junctions}
\label{sec5}

In sections \ref{sec:section_2} and \ref{sec:section_3}, we presented two approaches to compute the noise spectrum of a dc biased topological Josephson junction. While the range of validity of the two approaches is different, both are applicable in the regimes $R\ll1$, corresponding to $\delta\ll\Delta$, and $\tau_{\cal R}\to\infty$.

 In order to quantitatively compare the two approaches,  one may compute the switching probability $s$ of the phenomenological model starting from the Hamiltonian
\begin{equation}
\label{eq-Hfin}
\mathcal{H}=vp_x\sigma_z\tau_z+\Delta(x)e^{i\phi(x)\tau_z}\tau_x+M(x)\sigma_x-eU(x,t)\tau_z,
\end{equation}
see Eqs.~\eqref{eq:Hamiltonien_brut_stis_BdG} and \eqref{eq-HU}.

In the following, we consider the limit of a highly transmitting junction, where the separation between the bound state and the continuum is much smaller than the gap, $\delta\approx R\Delta/2\ll\Delta$, with the reflection probability $R=1-D\approx(ML/v)^2 \ll 1$. Furthermore, we restrict our attention to small bias, $eV_{\rm dc}\ll\Delta$.
Due to the applied bias, the superconducting phase acquires a finite phase velocity $\dot\varphi=2eV_{\rm{dc}}$ which enables non-adiabatic transitions between the bound state and the continuum. These non-adiabatic transitions occur in narrow phase intervals $|\varphi-2n\pi|\ll\pi$. To determine the transition probability $s$, we focus on the case $n=0$, corresponding to time intervals $|t|\ll\pi/(eV_{\rm{dc}})$.

As in section \ref{sec:section_3}, we use the unitary transformation $\mathcal U(t)=\exp{\left[i\phi(t)\tau_z/2\right]}$ to shift the time dependence from the reservoirs to the barrier. Taking the limit $L\to0$, while keeping $R$ fixed, the Hamiltonian (\ref{eq-Hfin}) becomes 
$\mathcal{H}=vp_x\sigma_z\tau_z+\Delta(x)\tau_x+v\left[ (\varphi/2)\sigma_z+\sqrt R\sigma_x\right]\delta(x)$. 
Furthermore, at $eV_{\rm{dc}}\ll\Delta$, only states close to the continuum edge, $v|p_x|\ll\Delta$, are relevant. Diagonalizing the bulk Hamiltonian
with a further unitary transformation $\mathcal W\approx \exp[-i\pi\sigma_z\tau_y/4]$, 
and restricting ourselves to the $2\times2$ subspace corresponding to positive energies (formed by the components $u_+$ and $v_-$ of the wave function), the reduced Hamiltonian reads
\begin{equation}\label{eq:Hamiltonien_effectif_stis_BdG}
\tilde{\mathcal H} = \Delta+\frac{(vp_x)^2}{2\Delta}+v\left[ \frac\varphi2\sigma_z+\sqrt R\sigma_x\right]\delta(x).
\end{equation}
The Hamiltonian $\tilde{\cal H}$ describes a spin degenerate continuum with quadratic dispersion, in the presence of a localized magnetic scatterer. In equilibrium, similarly to a magnetic impurity in a conventional superconductor~\cite{Yu1965,Shiba1968,Rusinov1969a,Rusinov1969b}, the magnetic barrier generates a localized bound state with energy $\epsilon_A(\varphi)=\Delta\left(1-R/2-\varphi^2/8\right)$
and wave function
\begin{equation}\label{eq:WF_ABSred}
\tilde\psi_A(x;\varphi)=\sqrt{\frac \Delta v}\left(R+\frac{\varphi^2}4\right)^{1/4}\exp\left[-i\frac\theta2 \sigma_y\right]
\left(\begin{array}{c} 0 \\ 1\end{array}\right)e^{-\kappa|x|},
\end{equation}
where $\theta=\arccos[\varphi/(2\sqrt{R+\varphi^2/4})]$ and $\kappa=(\Delta/v)\sqrt{R+\varphi^2/4}$, in accordance with Eqs.~(\ref{eq:ABS_WF_full})-\eqref{eq:WF_ABS2}
at $R,|\varphi|\ll1$.\footnote{
Indeed, at $R,|\varphi|\ll1$, Eqs.~\eqref{eq:ABS_WF_full}, \eqref{eq:WF_ABS}, and \eqref{eq:WF_ABS2} yield 
\begin{equation*}
\Psi_A(0)\approx\Psi_A(L)\approx \sqrt{\frac \Delta {2v}}\left(R+\frac{\varphi^2}4\right)^{1/4}\exp\left[-i\frac\theta2 \sigma_y\right]\begin{pmatrix}0&0&1&1\end{pmatrix}^T,
\end{equation*}
which coincides with Eq.~\eqref{eq:WF_ABSred} upon applying the rotation ${\cal W}^\dagger$, i.e.,
 \begin{equation*}
{\cal W}^\dagger\Psi_A(0)\approx \sqrt{\frac \Delta {v}}\left(R+\frac{\varphi^2}4\right)^{1/4} \begin{pmatrix} -\sin\frac\theta2&0&0&\cos\frac\theta2\end{pmatrix}^T,
\end{equation*}
and projecting on the subspace of states with positive energy.}
The wave functions $\tilde\psi_{p_x\pm}$ for the doubly degenerate states in
the continuum with energy $\epsilon=\Delta+(vp_x)^2/(2\Delta)$ may be found similarly.

At finite dc bias voltage, {$\varphi=2eV_{\rm dc}t$. Thus, the magnetic scatterer becomes time-dependent}. Due to the linear time dependence, the Hamiltonian \eqref{eq:Hamiltonien_effectif_stis_BdG} is a generalization to a two-band model of the problem of non-adiabatic transitions between a discrete state and a continuum, considered by Y. N. Demkov and V. I. Osherov~\cite{Demkov1968}.

\subsection{Bound state ionization rate}\label{subsubsec_212}

Near the continuum edge, within the time interval $|t|\ll\pi/(eV_{\rm{dc}})$, a particle occupying the bound state at $t\to-\infty$ has a probability $s$ to escape to the continuum as the phase increases. Dimensional analysis shows that the transition probability is governed by the adiabaticity parameter $\lambda=R^{3/2}\Delta/(eV_{\rm{dc}})$. 
Indeed, we may rescale the space and time coordinates by characteristic length and time scales, $\ell=v/\left[\Delta^2eV_{\rm{dc}}\right]^{1/3}$ and $\tau=1/\left[\Delta(eV_{\rm{dc}})^2\right]^{1/3}$, respectively. Then we find that the Schr\"odinger equation determined by the Hamiltonian (\ref{eq:Hamiltonien_effectif_stis_BdG}),
\begin{equation}
\label{eq:Schrod}
i\frac{\partial}{\partial t}\psi(x,t)=
\left[-\frac{1}2\partial_x^2+\left(t\sigma_z+\lambda^{1/3}\sigma_x\right)\delta(x)\right]\psi(x,t),
\end{equation}
where the wave function $\psi(x,t)$ is a two-component spinor, only depends on the parameter $\lambda$. Below we solve Eq.~\eqref{eq:Schrod} in various regimes to obtain the dependence of the switching probability on $\lambda$.

{
In the quasi-adiabatic limit, $\lambda\gg1$, it is convenient to express the exact wave function in the adiabatic basis of Eq.~(\ref{eq:Schrod}),
\begin{equation}
\psi(x,t)=c_A(t)e^{-i\int_0^t ds\varepsilon_A(s)}\psi_A(x,t)
+\sum_{p,\pm}c_{p\pm}(t)e^{-ip^2 t/2}\psi_{p\pm}(x,t),
\end{equation}
in terms of the amplitudes $c_A$ and $c_{p\pm}$ associated with the adiabatic wave functions for the Andreev bound state and the continuum states, respectively,
where, after rescaling, $\varepsilon_A(t)=\tau[\epsilon_A(\varphi(\tau t))-\Delta]$ and 
$\psi_{A/p\pm}(x,t)=\sqrt \ell\, \tilde\psi_{A/(\ell^{-1}p)\pm}(\ell x;\varphi(\tau t))$.

Using Eq.~\eqref{eq:WF_ABSred}, the adiabatic wave function of the Andreev bound state is given as
\begin{equation}
\label{eq:psiA}
\psi_A(x,t)=[-2\varepsilon_A(t)]^{1/4}
e^{-i\frac{\theta(t)}2\sigma_y}\left(\begin{array}{c}0\\1\end{array}\right)e^{-[-2\varepsilon_A(t)]^{1/2}|x|},
\end{equation}
where $\varepsilon_A(t)=-(t^2+\lambda^{2/3})/2$ and $\theta(t)=\arccos\left(t/[-2\varepsilon_A(t)]^{1/2}\right)$.

Since the bound state wave function as well as the time-dependent perturbation are even in $x$, the bound state dynamically couples only to the even wave functions of the continuum which, at energies $p^2/2$,  are given as
\begin{equation}
\psi_{p+}(t)=\sqrt{ 2 } \cos \left[p(|x|-x_0)\right]e^{-i\frac{\theta(t)}2\sigma_y}
\left(\begin{array}{c}1\\0\end{array}\right),
\qquad
\psi_{p-}(t)=\sqrt{ 2 } \cos \left[p(|x|+x_0)\right]e^{-i\frac{\theta(t)}2\sigma_y}
\left(\begin{array}{c}0\\1\end{array}\right),
\end{equation}
where $x_0=\arctan\left([-2\varepsilon_A(t)]^{1/2}/p\right)$ and $p>0$.

Using the initial conditions $c_A(-\infty)=1$ and $c_{p\pm}(-\infty)=0$, the switching probability corresponds to the probability at $t\to\infty$  to populate the continuum states, $s=\sum_{p\pm}\left|c_{p\pm}(\infty)\right|^2$. 

Writing the Schr\"odinger equation \eqref{eq:Schrod} in the adiabatic basis, we find that, in the considered quasi-adiabatic limit, $\psi_A$  couples with the combination $\psi_{p}=\sin\theta\,\psi_{p+}+\cos\theta\,\psi_{p-}$. 
The associated probability amplitudes $c_{p}(t)$ can be obtained using
\begin{equation}
\dot c_{p}(t)
=-i\frac{\int dx\; \psi^\dagger_{p}(x,t) \delta(x) \sigma_z \psi_A(x,t)}{p^2/2-\varepsilon_A(t)}e^{i\int^t{ds\,\left[p^2/2-\varepsilon_A(s)\right]}}.
\end{equation}
In particular, using $\varepsilon_A(t)=-(t^2+\lambda^{2/3})/2$, we find 
\begin{equation}
c_{p}(\infty)
=-i2\sqrt{2}p\int dt \frac{(t^2+\lambda^{2/3})^{1/4}}{[p^2+t^2+\lambda^{2/3}]^{3/2}}e^{i(p^2+\lambda^{2/3})t/2+it^3/6},
\end{equation}
and, thus,
\begin{equation}
\label{eq:sadiab}
s
=\frac 4\pi \int_0^\infty dp \;p^2\left|\int dt \frac{(t^2+\lambda^{2/3})^{1/4}}{[p^2+t^2+\lambda^{2/3}]^{3/2}}e^{i(p^2+\lambda^{2/3})t/2+it^3/6}\right|^2.
\end{equation}
Eq.~\eqref{eq:sadiab} can be evaluated using a saddle point method, where the saddle point in time is given as $t_0(p)=i\sqrt{p^2+\lambda^{2/3}}$.
Introducing new variables $t=t_0(p)+\sqrt{2}z/ (p^2+\lambda^{2/3})^{1/4}$ and $p= q/\lambda^{1/6}$, and recognizing that only variables $q,|z|\lesssim 1$ contribute to the integral~\eqref{eq:sadiab}, it can be simplified to
\begin{equation}
\label{eq:sadiab2}
s
=\frac {1}{2^{3/4}\pi} 
\lambda^{-5/4}e^{-2\lambda/3}
\int_0^\infty dq \;q^2{e^{-q^2}}
\left|\int dz \frac1{z^{5/4}}e^{-z^2}\right|^2,
\end{equation}
where the integration contour in $z$-plane should be chosen so that the corresponding integral is regular.
Evaluating the integrals in Eq.~\eqref{eq:sadiab2}, we obtain the switching probability $s\simeq {C_{\rm{q}}} \lambda^{-5/4}e^{-2\lambda/3}$ with  ${C_{\rm{q}}}=2^{13/4}\pi^{3/2}/ \Gamma^2(1/8)\simeq 0.93$.
The characteristic time scale for the transition is identified from the value of the saddle point in time, $\tau_{\rm t}\sim |t_0(0)|\tau \sim\sqrt R/(eV_{\rm{dc}})$.

Let us now consider the opposite, anti-adiabatic limit, $\lambda\ll 1$.}
At $\lambda=0$ the spin bands are decoupled. At times $t<0$, the bound state belongs to the spin-up band whereas, at times $t>0$, it belongs to the spin-down band. The spin-up band is described by the wave function
\begin{equation}
\label{eq:psi+}
\psi_+(x,t)=f(x,t)
\left(\begin{array}{c}1\\0\end{array}\right),
\end{equation} 
while the spin-down band is obtained by time reversal,  
\begin{equation}
\label{eq:psi-}
\psi_-(x,t)=f(x,-t)^*
\left(\begin{array}{c}0\\1\end{array}\right).
\end{equation} 
To determine $f(x,t)$, we follow the method described in Ref.~\cite{Demkov1968} and introduce 
\begin{equation}
f(x,t)=\int_{-\infty}^\infty\frac{dp}{2\pi}\int_{\mathcal C} d\omega f(p,\omega)e^{ip x-i\omega t}, 
\end{equation}
where $\mathcal C$ is an infinite contour in the complex $\omega$-plane, to be specified below. Then Eq.~\eqref{eq:Schrod} yields
\begin{equation}
\label{eq:Schrod2}
\left(-\omega+\frac{p^2}2\right)f(p,\omega)=i\frac{\partial}{\partial \omega}\int\frac{dp}{2\pi}f(p,\omega).
\end{equation}
{Dividing both sides of Eq.~\eqref{eq:Schrod2} by $p^2/2-\omega$ and summing over $p$, we obtain a differential equation 
for $f(x=0,\omega)=\int dp/(2\pi)  f(x=0,\omega)$. It is solved with 
\begin{equation}
\label{eq:solf}
f(x=0,\omega)={\cal N} \exp\left[\frac i3(-2\omega)^{3/2}\right],
\end{equation}
provided that the contour $\mathcal C$ starts and ends at infinity, with arguments $\pi<\theta<5\pi/3$ and $0<\theta<\pi/3$, respectively, and avoids a branch cut along the positive real axis, cf.~Fig.~\ref{fig:Fig-contour}. The normalization factor ${\cal N}$ can be obtained by realizing that, in the limit $t\to-\infty$, the wave function $\psi_+(x,t)$ should coincide with the adiabatic wave function of the bound state, $\psi_A(t)\exp[-i\int^t ds\;\varepsilon_A(s)]$. In that limit, the function $f(0,t)$ can be evaluated using a saddle point method to find 
$f(x=0,t\to-\infty)\approx{\cal N}\sqrt{-2\pi t}\exp[i\pi/4+it^3/6]$.
Comparison with Eq.~\eqref{eq:psiA} using $\varepsilon_A(t)=-t^2/2$ then yields ${\cal N}=1/\sqrt{2\pi}$ up to an unimportant phase factor.}
\begin{figure}[ht!]
\begin{center}
\includegraphics[scale=0.4]{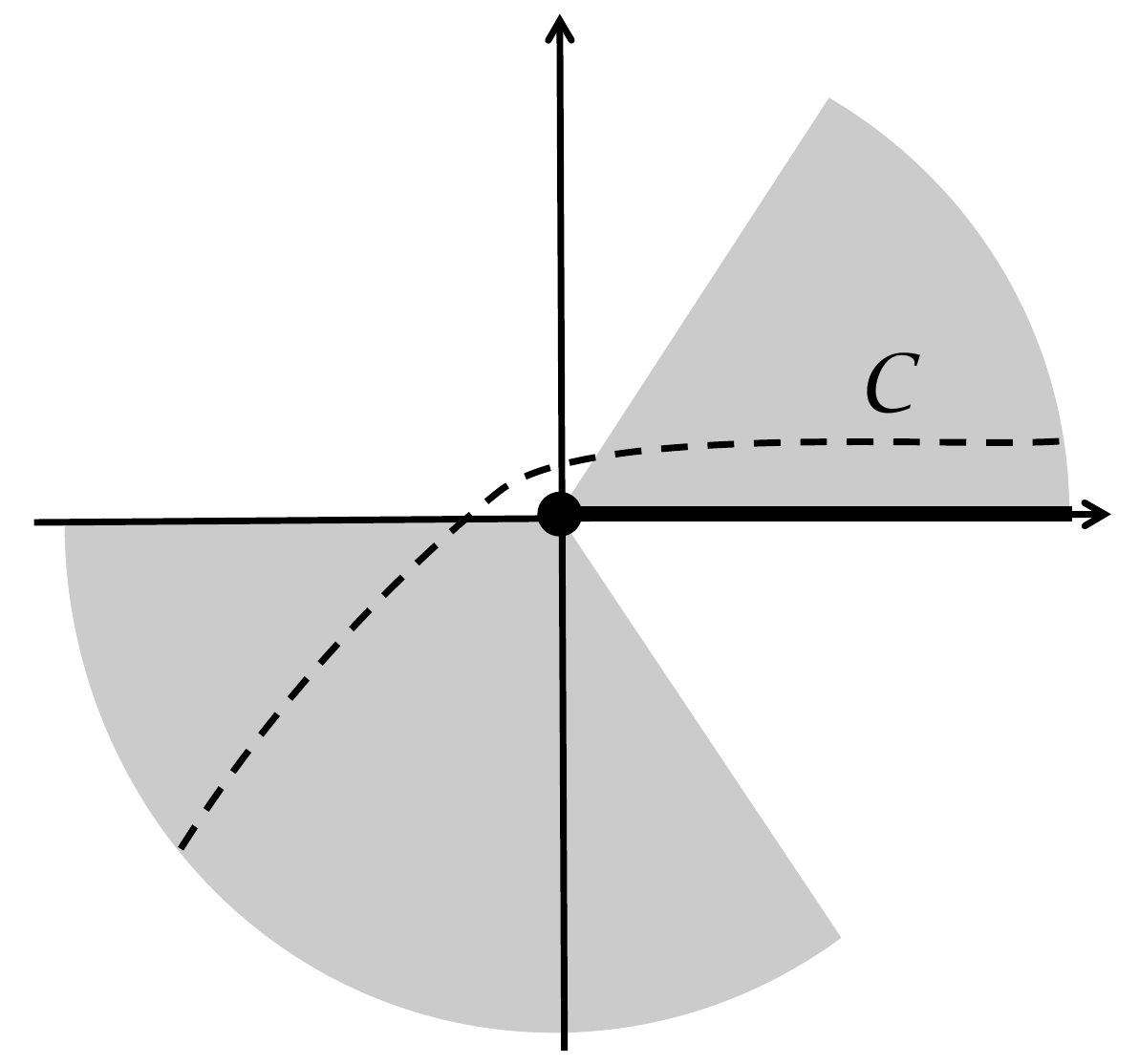}
\caption{Integration contour $\mathcal{C}$ in the complex $\omega$-plane. The contour has to begin and end within the shaded areas, where  $f(x=0,\omega)$ vanishes in the limit $|\omega|\to\infty$, and avoid the branch cut along the positive real axis.}
\label{fig:Fig-contour}
\end{center}
\end{figure}

At a finite value of the adiabaticity parameter $\lambda$, the two spin bands are coupled, thus spin flips may occur. 
{
The switching probability $s$ is obtained from the overlap of the exact wave function $\psi$ with the spin-down wave function,
$s=1-\left|\int dx \;\psi_-^\dagger(x,\infty)\psi(x,\infty)\right|^2.$
At $t=-\infty$, the exact wave function is given by the spin-up bound state, $\psi(x,-\infty)=\psi_+(x,-\infty)$. 
Looking for an exact wave function in the form $\psi(x,t)=\psi_+(x,t)+\delta\psi(x,t)$ and using Eq.~\eqref{eq:Schrod}, one obtains 
\begin{equation}
\label{eq:perturb}
\delta\psi(x,t)\approx -i\int_{-\infty}^tds\;\lambda^{1/3}\delta(x)\sigma_x\psi_+(x,s),
\end{equation}
perturbatively in $\lambda\ll1$.
Using Eqs.~\eqref{eq:psi+} and \eqref{eq:psi-} for the wave functions $\psi_\pm$ and the fact that $\psi_+$ and $\psi_-$ are orthogonal, the overlap is thus obtained through
\begin{equation}
\int dx\; \psi_-^\dagger(x,\infty)\psi(x,\infty)
=
-i\lambda^{1/3}
\int^\infty_{-\infty} ds
\int dx \;f(x,-s)\delta(x)f(x,s)
=-i\lambda^{1/3}\int_{\mathcal C}d\omega\; e^{i2(-2\omega)^{3/2}/3}.
\end{equation}
}Computing the transition probability, we find $s\approx1-{C_{\rm{a}}}\lambda^{2/3}$, where ${C_{\rm{a}}}=3^{1/3}2^{-4/3}\Gamma^2(2/3)\approx 1.05$.
The typical time scale for the transition is $\tau_{\rm t}\sim\tau$.

Note that in both limits, $\lambda\gg1$ and $\lambda\ll1$, we find that the characteristic time scale for the transition $\tau_{\rm t}$ is much smaller than the Josephson oscillation period. This justifies the assumption used for the phenomenological model in Sec.~\ref{sec:section_2} that switching takes place in narrow phase intervals around $\varphi=2n\pi$.

For an arbitrary adiabaticity parameter $\lambda$, the transition amplitude can be obtained numerically by discretizing Eq. (\ref{eq:Hamiltonien_effectif_stis_BdG}) on a tight binding lattice. The computed switching probability, with the asymptotes obtained above, are shown in Fig. \ref{fig:Fig02}.

\begin{figure}[ht!]
\begin{center}
\includegraphics[scale=0.6]{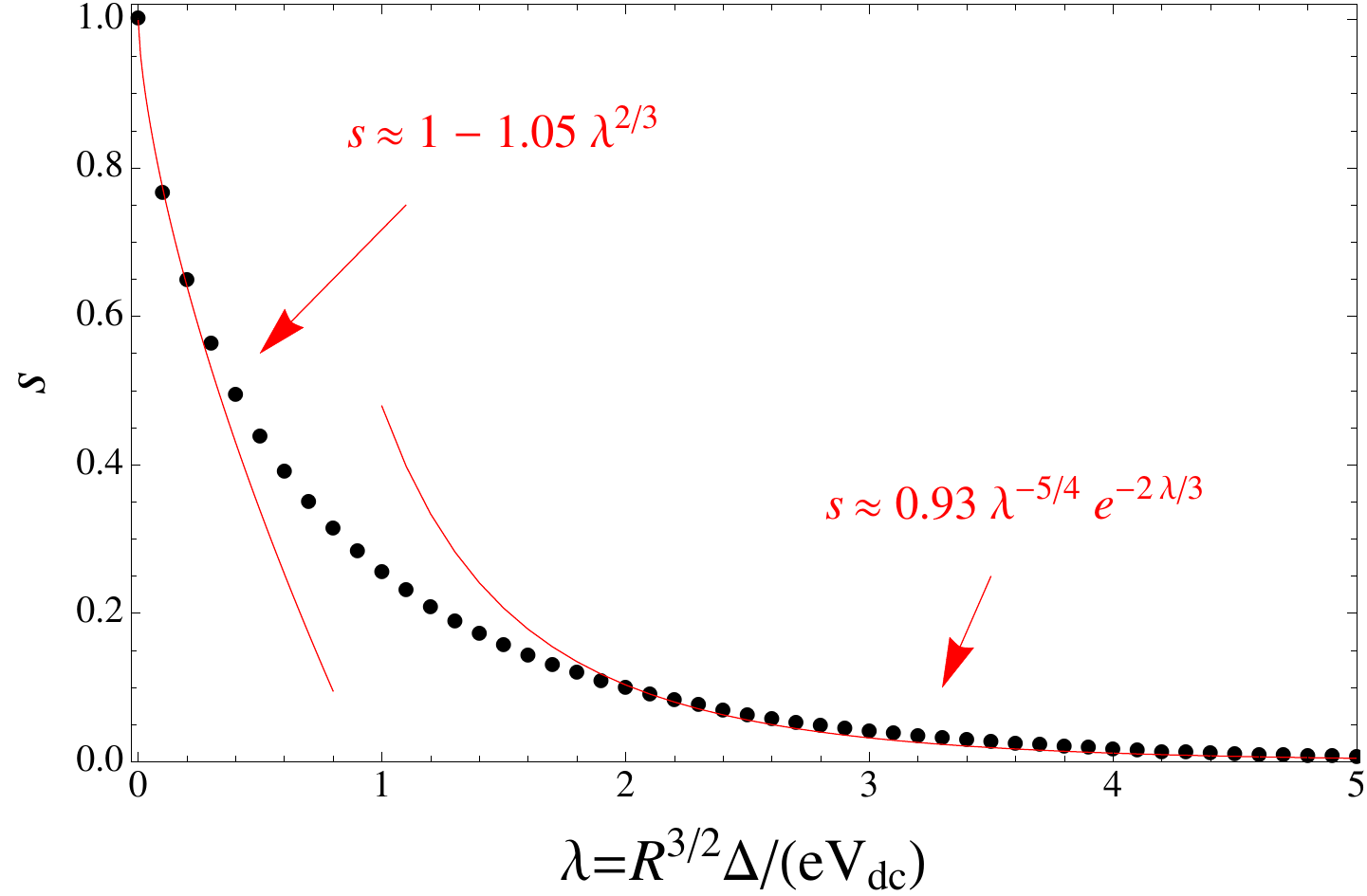}
\caption{Switching probability $s$ as a function of the adiabaticity parameter $\lambda$. Dots: $s$ found by solving Eq. (\ref{eq:Hamiltonien_effectif_stis_BdG}) numerically. Lines: Asymptotes obtained analytically in the anti-adiabatic ($\lambda\ll1$) and quasi-adiabatic ($\lambda\gg1$) limits.}
\label{fig:Fig02}
\end{center}
\end{figure}

\subsection{Comparison of Multiple-Andreev-Reflection and Demkov-Osherov approaches}

Having obtained the switching probability as a function of the junction parameters, we are now in a position to compare the results of Secs. \ref{sec:section_2} and \ref{sec:section_3} in the regimes $R\ll1$, corresponding to $\delta\ll\Delta$, and $\tau_{\cal R}\to\infty$. In particular, we may compare the numerical curves for the noise spectrum with the analytic result, Eq.~\eqref{eq:Svsomega}. At fixed transmission and voltage, we use $s$ as a fitting parameter to fit the numerical curves. In a next step we then compare the fit parameter with the switching probability as function of transmission and voltage obtained in the previous section. The results are shown in Fig.~\ref{fig:figure5}. The agreement is excellent in all regimes.

\begin{figure}[ht!]
\centering
\includegraphics[scale=0.6]{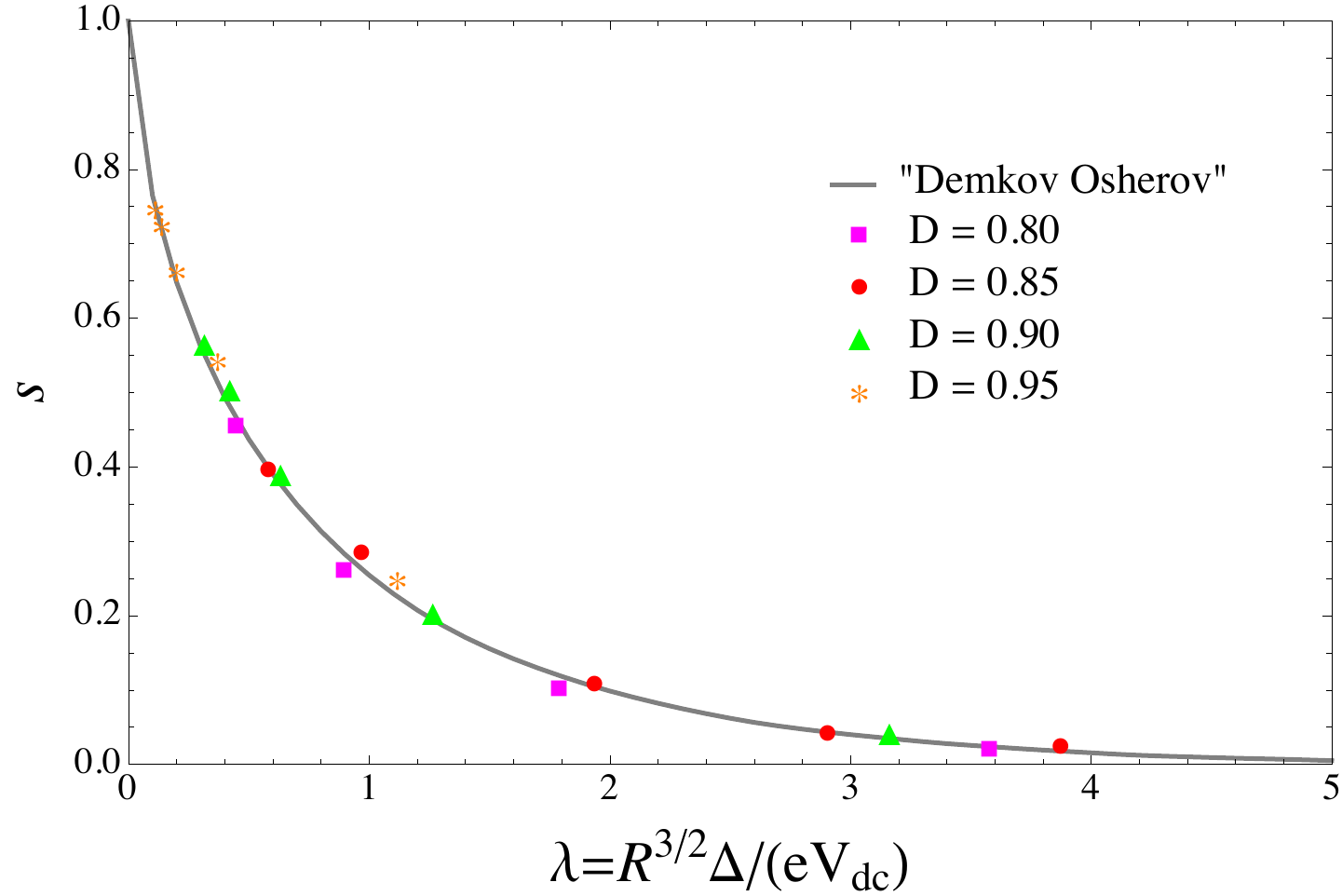}
\caption{Switching probability $s$ as a function of the adiabaticity parameter $\lambda$. Comparison of the fit to the MAR results for different transmissions (symbols) with the microscopic values (line) as shown in Fig.~\ref{fig:Fig02}. 
}
\label{fig:figure5}
\end{figure}

\section{Signatures of the midgap states in the stationary current}\label{sec6}

So far we concentrated on probing the fractional Josephson effect. It turns out, however, that this is not the only signature of the presence of Majorana fermions in biased topological Josephson junctions. Namely the presence of the zero-energy bound state also shows up in the current-voltage characteristic of a dc biased junction~\cite{Badiane2011,Aguado}. 
{
Previously, this effect had been predicted  for Josephson junctions formed with $d$-wave superconductors having specific orientations of their crystallographic axes with respect to the interface~\cite{mar0,mar1}, as well as for Josephson junctions where the superconducting leads are coupled through a magnetic impurity \cite{mar2}
}.

Using the formalism presented in Sec.~\ref{sec:section_3}, we may compute the dc current,
\begin{equation}
I_0=
\frac eh
\left\{
DeV_{\rm dc}
-\int d\epsilon\, \tanh\frac{\epsilon}{2T}
J^2
\left[2a_0\Re[A_0]
+
\sum_m(1+|a_{2m}|^2)
\left(|A_{m}|^2-|B_{m}|^2\right)
\right]\right\}.
\label{eq:dc-current}
\end{equation}
The results for different transmission probabilities of the junction are shown in Fig.~\ref{fig:Fig03}.

\begin{figure}[ht!]
\begin{center}
\includegraphics[scale=0.2]{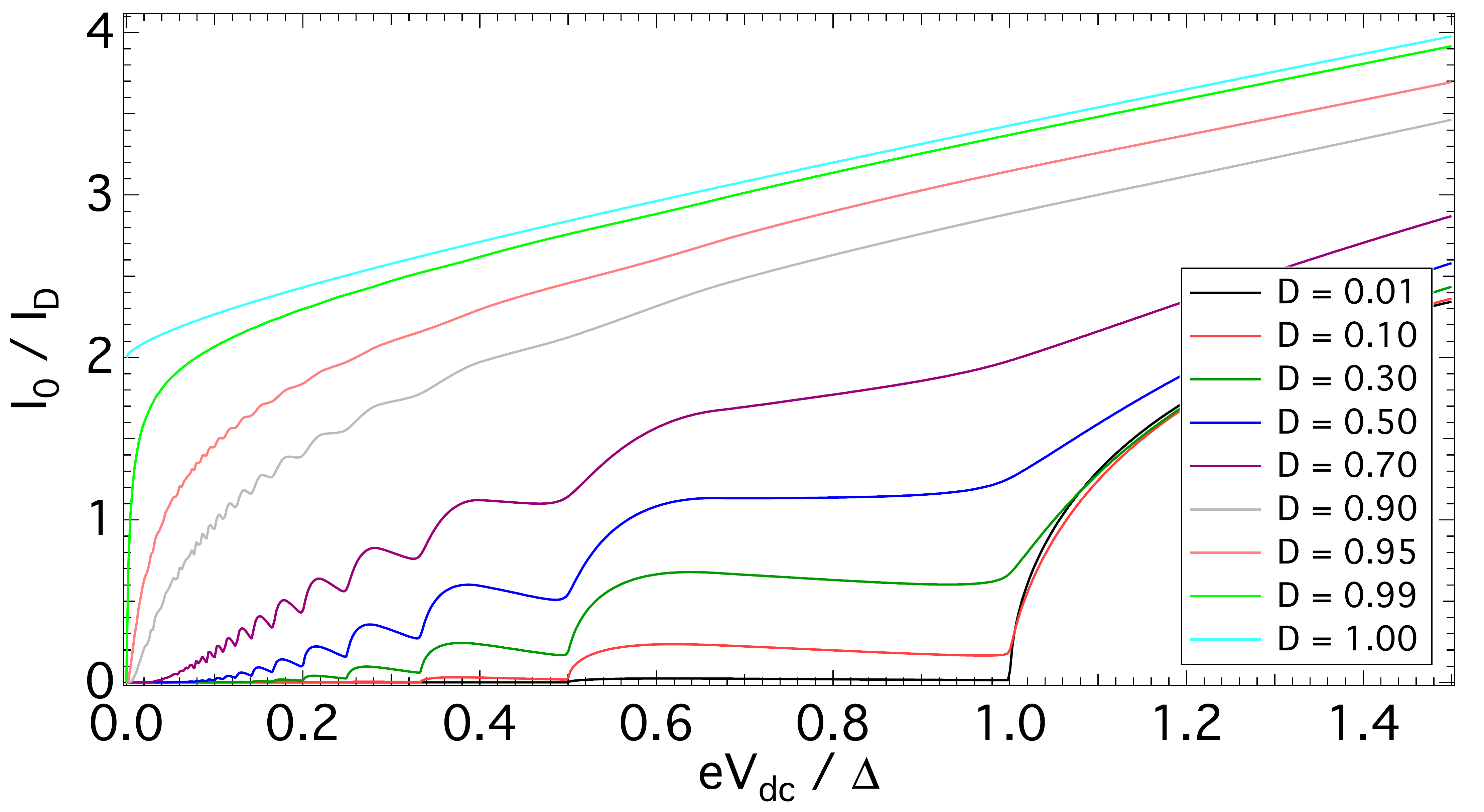}
\caption{Dc current $I_0$ as a functions of applied voltage for different values of the transmission. Multiple Andreev reflection signatures appear at voltages $eV_{\rm dc}=\Delta/n$.  
}
\label{fig:Fig03}
\end{center}
\end{figure}

Most striking is the curve in the tunneling limit ($D=0.01$), where we see a current onset at $eV_{\rm dc}=\Delta$. This has to be contrasted with the I-V characteristics of a conventional Josephson junction where the current onset happens at $eV_{\rm dc}=2\Delta$. The current onset at $eV_{\rm dc}=\Delta$ can be attributed to the presence of a midgap bound state. Namely a quasi-particle injected from the continuum of filled states below the gap needs to gain an energy $\Delta$ to reach the bound state. By contrast, in the conventional case, transitions are possible only between the continuum of filled states below the gap and the continuum of empty states above the gap, thus involving an energy cost of at least $2\Delta$.

In the tunneling limit, an analytic expression for the current may be obtained using the tunnel Hamiltonian. Namely,
\begin{equation}
I_0^\mathrm{tun}(V_{\rm dc})=
\frac{eD}h
\int d\epsilon\,
\nu_L(\epsilon)\nu_R(\epsilon-eV_{\rm dc})[f(\epsilon-eV_{\rm dc})-f(\epsilon)],
\end{equation}
where $\nu_{L/R}$ are the normalized local densities of states to the left and to the right of the junction, respectively. Furthermore, $f(\epsilon)$ is the Fermi distribution.

The local density of states can be computed from the wave functions at the barrier in the limit $D\to0$, found in Sec.~\ref{sec:section_1} for the bound state and obtained by a generalization of the results of Sec.~\ref{sec5} for the continuum. One finds
\begin{equation}
\nu_{L/R}(\epsilon)
=
\pi \Delta \delta(\epsilon)+\theta\left(\epsilon^2-\Delta^2\right)\sqrt{1-\left(\frac\Delta\epsilon\right)^2},
\end{equation}
which shows the contributions of the bound state at $\epsilon=0$ and of the continuum at $|\epsilon|>\Delta$.

As a consequence, at $T=0$, the current 
\begin{equation}
I_0^\mathrm{tun}=\frac{eD}h\left\{\theta(eV_{\rm dc}-\Delta)\pi\Delta\sqrt{1-\left(\frac{\Delta}{eV_{\rm dc}}\right)^2}+\theta(eV_{\rm dc}-2\Delta)\int_\Delta^{eV_{\rm dc}-\Delta}d\epsilon\;\sqrt{1-\left(\frac{\Delta}{\epsilon}\right)^2}\sqrt{1-\left(\frac{\Delta}{eV_{\rm dc}\!-\!\epsilon}\right)^2}\right\}
\end{equation}
 is the sum of two terms. The first term corresponds to the transitions between the continuum and the bound state for voltages $eV_{\rm dc}\geq\Delta$ and is, thus, due to the presence of Majorana fermions in the junction. By contrast,  the second term corresponds to the transitions from continuum to continuum for voltages $eV_{\rm dc}\geq2\Delta$ and is present in conventional junctions as well. Due to the suppression of the BCS square-root singularity in the local density of states $\nu_{L/R}(\epsilon)$, the singular behavior of $I_0(V_{\rm dc})$ at $eV_{\rm dc}=2\Delta$ is smooth (see Fig.~\ref{fig:Fig03}). 

At higher transmissions, multiple Andreev reflections lead to non-analyticities in the I-V characteristics at $neV_{\rm dc}=\Delta$ when a new channel for transport opens between the continuum and the bound state. The non-analyticities at $meV_{\rm dc}=2\Delta$ corresponding to transitions between the continuum states below and above the gap are present as well, but weaker than in the conventional case due to the modified density of states, as discussed above.

The MAR signatures at voltages $eV_{\rm dc}=\Delta/n$ thus provide a clear signature of the midgap bound state due to the presence of Majorana fermions in a topological Josephson junction~\cite{Badiane2011}. Recently this effect has been studied in detail~\cite{Aguado} for topological Josephson junctions based on nanowires with strong spin-orbit coupling, where the transition between a topologically trivial and a topologically non-trivial phase can be tuned by an applied Zeeman field $B_Z$. In that case, for $\mu=0$, the MAR features are expected at $eV_{\rm dc}=2(\Delta-B_Z)/n$ on the topologically trivial side and at $eV_{\rm dc}=(B_Z-\Delta)/n$ on the topologically non-trivial side of the transition.

\section{Conclusion}\label{sec:section_4}

The prospect of realizing Majorana fermions in superconducting hybrid systems has led to considerable excitement in the community. While possible experimental signatures have been reported in the last two years, more studies are necessary. In this context, we studied the non-equilibrium properties of topological Josephson junctions. In particular, we discussed the observability of the fractional Josephson effect  as well as the presence of characteristic MAR features in the current-voltage characteristics associated with the  Majorana fermions.

The observability of the fractional Josephson effect depends on two characteristic time scales, namely the lifetime of the Andreev bound state $\tau_{\rm s}$, which is necessarily finite due to its dynamical coupling with the continuum,  and the phase adjustment time across the junction $\tau_{\mathcal R}$, due to the external circuit. We showed that the fractional Josephson effect manifests itself in an even-odd effect in the Shapiro steps when $\tau_{\rm s}\gg\tau_{\cal R}$, namely in that case only the even Shapiro steps are visible. In the opposite limit, $\tau_{\rm s}\ll\tau_{\cal R}$, the fast switching of the occupation of the bound state suppresses all Shapiro steps. In this regime, signatures of the fractional Josephson effect nevertheless survive in the finite-frequency current noise. In particular, the noise spectrum $S(\omega)$ displays a peak at $\omega=eV_{\rm dc}$ whose width is determined by $1/\tau_{\rm s}$.

Note that the critical current in a short Josephson junction, as considered here, does not depend on the occupation of the bound state. Thus, the switching current under dc bias is the same as the maximal current obtained from the equilibrium Josephson relation. However, it has been pointed out recently~\cite{Beenakker-long} that, in long ballistic junctions, the switching current for $\tau_{\rm s}\gg\tau_{\cal R}$ differs from the maximal equilibrium current by a factor of 2. Thus, for those junctions, a comparison between the switching current and the maximal equilibrium current may give additional evidence of the $4\pi$-periodicity at fixed parity. 

Finally, the dissipative dc current provides further signatures of the presence of Majorana fermions. The MAR features of the I-V characteristics in the subgap regime are associated with the opening of additional current channels across the junction. In the presence of a zero-energy bound state, this happens when $eV_{\rm dc}=\Delta/n$ ($n\in\mathbb{N}$). By contrast, in conventional junctions, these channels are associated with transitions from the continuum of filled states below the gap to the continuum of empty states above the gap and are thus determined by the condition $eV_{\rm dc}=2\Delta/n$.

To summarize, while recent experiments are promising, further signatures are necessary to confirm the realization of Majorana bound states in proximity-based topological superconductors. Voltage-biased topological Josephson junctions provide several such signatures which hopefully will be explored in the near future.

\section*{Acknowledgements}

Work at the INAC/SPSMS was supported through ANR Grants No. ANR-11-JS04-003-01 and No. ANR-12- BS04-0016-03, and an EU-FP7 Marie Curie IRG. Work at Yale University was supported by DOE BES under Contract No. DEFG02-08ER46482.

\appendix


\begin{thebibliography}{00}

\bibitem{MajoranaEttore1937} E. Majorana, Nuovo Cimento \textbf{14}, 171 (1937).

\bibitem{Alicea2012} J. Alicea, Rep. Prog. Phys. \textbf{75}, 076501 (2012).

\bibitem{rev-flensberg} M. Leijnse and K. Flensberg, Semicond. Sci. Technol. \textbf{27}, 124003 (2012).

\bibitem{rev-beenakker} C. W. J. Beenakker, Annu. Rev. Condens. Matter Phys. \textbf{4}, 113 (2013).

\bibitem{Kitaev2003} A. Y. Kitaev, Ann. Phys. (N.Y.) \textbf{303}, 2 (2003).

\bibitem{Bravyi2002} S. B. Bravyi and A. Y. Kitaev, Ann. Phys. (N.Y.) \textbf{298}, 210 (2002).

\bibitem{MooreRead1991} G. Moore and N. Read, Nucl. Phys. B \textbf{360}, 362 (1991).

\bibitem{Kopnin1991} N. B. Kopnin and M. M. Salomaa, Phys. Rev. B \textbf{44}, 9667 (1991).

\bibitem{Levitov2001} L. S. Levitov, T. P. Orlando, J. B. Majer, and J. E. Mooij, arXiv:cond-mat/0108266 (2001).

\bibitem{Kitaev2001} A. Y. Kitaev, Phys. Usp. \textbf{44}, 131 (2001).

\bibitem{Volovik1999} G. E. Volovik, JETP Lett. \textbf{70}, 609 (1999).

\bibitem{ReadGreen2010} N. Read and D. Green, Phys. Rev. B \textbf{61}, 10267 (2000)

\bibitem{Ivanov2001} D. A. Ivanov, Phys. Rev. Lett. \textbf{86}, 268 (2001).

\bibitem{Mackenzie2003} A. P. Mackenzie and Y. Maeno, Rev. Mod. Phys. \textbf{75}, 657, (2003).

\bibitem{Fu2008} L. Fu and C. Kane, Phys. Rev. Lett. \textbf{100}, 096407 (2008).

\bibitem{Fu2009} L. Fu and C. Kane, Phys. Rev. B \textbf{79}, 161408(R) (2009).

\bibitem{Bernevig2006} B. A. Bernevig, T. L. Hughes, and S.-C. Zhang, Science \textbf{314}, 1757 (2006).

\bibitem{Liu2008} C. Liu, T. Hughes, X.-L. Qi, K. Wang, and S.-C. Zhang, Phys. Rev. Lett. \textbf{100}, 236601 (2008).

\bibitem{Hasan2010} M. Hasan and C. Kane, Rev. Mod. Phys. \textbf{82}, 3045 (2010).

\bibitem{Qi2011} X.-L. Qi and S.-C. Zhang, Rev. Mod. Phys. \textbf{83}, 1057 (2011).

\bibitem{Konig2007} M. Konig, S. Wiedmann, C. Brune, A. Roth, H. Buhmann, L. W. Molenkamp, X.-L. Qi, and S.-C. Zhang, Science \textbf{318}, 766 (2007).

\bibitem{Knez2011} I. Knez, R. R. Du, and G. Sullivan, Phys. Rev. Lett. \textbf{107}, 136603 (2011).

\bibitem{Lutchyn2010} R. Lutchyn, J. Sau, and S. Das Sarma, Phys. Rev. Lett. \textbf{105}, 077001 (2010).

\bibitem{Oreg2010} Y. Oreg, G. Refael, and F. von Oppen, Phys. Rev. Lett. \textbf{105}, 177002 (2010).

\bibitem{Law2009} K. T. Law, P. A. Lee, and T. K. Ng, Phys. Rev. Lett. \textbf{103}, 237001 (2009).

\bibitem{Fidkowski2012} L. Fidkowski, J. Alicea, N. H. Lindner, R. M. Lutchyn, and M. P. A. Fisher, Phys. Rev. B \textbf{85}, 245121 (2012).

\bibitem{Mourik2012} V. Mourik, K. Zuo, S. M. Frolov, S. R. Plissard, E. P. A. M. Bakkers, and L. P. Kouwenhoven, Science \textbf{336}, 1003 (2012).

\bibitem{Das2012} A. Das, Y. Ronen, Y. Most, Y. Oreg, M. Heiblum, and H. Shtrikman, Nat. Phys. \textbf{8}, 887 (2012).

\bibitem{Liu2012} Jie Liu, A. C. Potter, K. T. Law, and P. A. Lee, Phys. Rev. Lett. \textbf{109}, 267002 (2012).

\bibitem{bagrets2012} D. Bagrets and A. Altland, Phys. Rev. Lett. \textbf{109}, 227005 (2012).

\bibitem{Lee2012} E. J. H. Lee, X. Jiang, R. Aguado, G. Katsaros, C. M. Lieber, and S. D. Franceschi,  Phys. Rev. Lett. \textbf{109}, 186802 (2012).

\bibitem{Lee2013} E. J. H. Lee, X. Jiang, M. Houzet, R. Aguado, C. M. Lieber, and S. D. Franceschi,  arXiv:1302.2611 (2013).

\bibitem{Kwon2004} H.-J. Kwon, K. Sengupta, and V. M. Yakovenko, Eur. Phys. J. B \textbf{37}, 349 (2003).

\bibitem{Shapiro1963} S. Shapiro, Phys. Rev. Lett. \textbf{11}, 80 (1963).

\bibitem{Liang2011} L. Jiang, D. Pekker, J. Alicea, G. Refael, Y. Oreg, and F. von Oppen, Phys. Rev. Lett. \textbf{107}, 236401 (2011).

\bibitem{Dominguez2012} F. Dominguez, F. Hassler, and G. Platero, Phys. Rev. B \textbf{86}, 140503 (2012).

\bibitem{Yanson} I. K. Yanson, V. M. Svistunov, and I. M. Dmitrenko, Sov. Phys. JETP \textbf{21}, 650  (1965).

\bibitem{Yanson2} I. M. Dmitrenko and I. K. Yanson, JETP Lett. \textbf{2}, 154 (1965).

\bibitem{Badiane2011} D. M. Badiane, M. Houzet, and J. S. Meyer, Phys. Rev. Lett. \textbf{107}, 177002 (2011).

\bibitem{Ivanchenko1969} Y. M. Ivanchenko and L. A. Zil'berman, Sov. Phys. JETP \textbf{28}, 1272 (1969).

\bibitem{Ambegaokar1969} V. Ambegaokar and B. I. Halperin, Phys. Rev. Lett. \textbf{22}, 1364 (1969).

\bibitem{Anderson1969} J. Anderson and A. Goldman, Phys. Rev. Lett. \textbf{23}, 128 (1969).

\bibitem{Houzet2013} M. Houzet, J. S. Meyer, D. M. Badiane, and L. I. Glazman, Phys. Rev. Lett. \textbf{111}, 046401 (2013).

\bibitem{noise-sns1} D. Averin and H. T. Imam, Phys. Rev. Lett. \textbf{76}, 3814 (1996).

\bibitem{noise-sns2} A. Mart\'{i}n-Rodero, A. Levy Yeyati, and F. J. Garc\'{i}a-Vidal, Phys. Rev. B \textbf{53}, 8891(R) (1996).

\bibitem{RainisLoss2012} D. Rainis and D. Loss, Phys. Rev. B \textbf{85}, 174533 (2012).

\bibitem{Aguado-transients} P. San-Jose, E. Prada, and R. Aguado, Phys. Rev. Lett. \textbf{108}, 257001 (2012).

\bibitem{Nazarov-noise} D. I. Pikulin and Y. V. Nazarov, Phys. Rev. B \textbf{86}, 140504(R) (2012)

\bibitem{Sau2012} J. D. Sau, E. Berg, and B. I. Halperin, arXiv:1206.4596 (2012).

\bibitem{virtanen-recher} P. Virtanen and P. Recher, arXiv:1303.2353 (2013).

\bibitem{S-quantronics} M. Chauvin, {\em The Josephson Effect in Atomic Contacts} (Ph.D. thesis, Universit\'{e} Paris 6, 2005).

\bibitem{Rokhinson2012} L. P. Rokhinson, X. Liu, and J. K. Furdyna, Nat. Phys. \textbf{8}, 795 (2012).

\bibitem{exp-4pi} P.-M. Billangeon, F. Pierre, H. Bouchiat, and R. Deblock, Phys. Rev. Lett. {\bf 98}, 216802 (2007).

\bibitem{freq-noise} R. J. Schoelkopf, P. J. Burke, A. A. Kozhevnikov, D. E. Prober, and M. J. Rooks, Phys. Rev. Lett. \textbf{78}, 3370 (1997).

\bibitem{Deblock2003} R. Deblock, E. Onac, L. Gurevich, and L. P. Kouwenhoven, Science {\bf 301}, 203  (2003).

\bibitem{Averin1995} D. Averin and A. Bardas, Phys. Rev. Lett. \textbf{75}, 1831 (1995).

\bibitem{Bratus1995} E. N. Bratus, V. S. Shumeiko, and G. Wendin, Phys. Rev. Lett. \textbf{74}, 2110 (1995).

\bibitem{Cuevas1996} J. C. Cuevas, A. Mart\'in-Rodero, and A. L. Yeyati, Phys. Rev. B \textbf{54}, 7366 (1996).

\bibitem{Yu1965} L. Yu, Acta Phys. Sin. \textbf{21}, 75 (1965).

\bibitem{Shiba1968} H. Shiba, Prog. Theor. Phys. \textbf{40}, 435 (1968).

\bibitem{Rusinov1969a} A. I. Rusinov, Sov. Phys. JETP \textbf{29}, 1101 (1969).

\bibitem{Rusinov1969b} A. I. Rusinov, Sov. Phys. JETP Lett. \textbf{9}, 85 (1969).

\bibitem{Demkov1968} Y. N. Demkov and V. I. Osherov, Sov. Phys. JETP \textbf{26}, 916 (1968).

\bibitem{Aguado} P. San-Jose, J. Cayao, E. Prada, and R. Aguado, New J. Phys. \textbf{15}, 075019 (2013).

\bibitem{mar0} T. L\"ofwander, G. Johansson, and G. Wendin, J. Low. Temp. Phys. \textbf{117}, 593 (1999).

\bibitem{mar1} J. C. Cuevas and M. Fogelstr\"om, Phys. Rev. B \textbf{64}, 104502 (2001).

\bibitem{mar2} M. Andersson, J. C. Cuevas, and M. Fogelstr\"om, Physica C \textbf{367}, 117 (2002).

\bibitem{Beenakker-long} C. W. J. Beenakker, D. I. Pikulin, T. Hyart, H. Schomerus, and J. P. Dahlhaus, Phys. Rev. Lett. {\bf 110}, 017003 (2013).

\end{thebibliography}
\end{document}